\documentclass[12pt]{iopart}
\usepackage{iopams}
\usepackage{multirow}
\usepackage{hyperref}
\usepackage{graphicx}

\newcommand{\qmcpack}{\texttt{QMCPACK} }

\newcommand{\abs}[1]{| #1 |}

\newcommand{\ket}[1]{| #1 \bigr>}
\newcommand{\bra}[1]{\bigl< #1 |}
\newcommand{\expval}[3]{\bra{#1}#2\ket{#3}}

\newcommand{\operator}[3]{\ket{#1} #2 \bra{#3}}

\newcounter{authcounter}
\newcommand*{\authlabel}[1]{\refstepcounter{authcounter}\theauthcounter\label{#1}}
\newcommand*{\authref}[1]{\ref{#1}}

\begin{document}

\title[\qmcpack: An open source ab initio Quantum Monte Carlo package]{\qmcpack: An open source ab initio Quantum Monte Carlo package for the electronic structure of atoms, molecules, and solids}

\author{Jeongnim Kim$^{\authref{intel}}$,
Andrew Baczewski$^{\authref{hedp-snl}}$,
Todd D. Beaudet$^{\authref{eng-uva}}$,
Anouar Benali$^{\authref{alcf-anl},\authref{csd-anl}}$,
M~Chandler Bennett$^{\authref{phy-ncsu}}$,
Mark A Berrill$^{\authref{csmd-ornl}}$,
Nick S Blunt$^{\authref{chem-cam}}$,
Edgar Josu\'e Landinez Borda$^{\authref{llnl}}$,
Michele Casula$^{\authref{sorbonne-fr}}$,
David M Ceperley$^{\authref{phy-ui}}$,
Simone Chiesa$^{\authref{phy-ui}}$,
Bryan K Clark$^{\authref{phy-ui}}$,
Raymond C Clay III$^{\authref{hedp-snl}}$,
Kris T Delaney$^{\authref{mrl-uc}}$,
Mark Dewing$^{\authref{csd-anl}}$,
Kenneth P Esler$^{\authref{stonert}}$,
Hongxia Hao$^{\authref{chem-brown}}$,
Olle Heinonen$^{\authref{msd-anl},\authref{ise-nw}}$,
Paul R C Kent$^{\authref{cnms-ornl},\authref{csed-ornl}}$,
Jaron T Krogel$^{\authref{mstd-ornl}}$,
Ilkka Kyl\"anp\"a\"a$^{\authref{mstd-ornl}}$,
Ying Wai Li$^{\authref{nccs-ornl}}$,
M Graham Lopez$^{\authref{csmd-ornl}}$,
Ye Luo$^{\authref{alcf-anl},\authref{csd-anl}}$,
Fionn D. Malone$^{\authref{llnl}}$,
Richard M Martin$^{\authref{phy-ui}}$,
Amrita Mathuriya$^{\authref{intel}}$,
Jeremy McMinis$^{\authref{llnl}}$,
Cody A Melton$^{\authref{phy-ncsu}}$,
Lubos Mitas$^{\authref{phy-ncsu}}$,
Miguel A Morales$^{\authref{llnl}}$,
Eric Neuscamman$^{\authref{chem-uc},\authref{csd-lbnl}}$,
William D Parker$^{\authref{uw-parkside}}$,
Sergio D Pineda Flores$^{\authref{chem-uc}}$,
Nichols A Romero$^{\authref{alcf-anl},\authref{csd-anl}}$,
Brenda M Rubenstein$^{\authref{chem-brown}}$,
Jacqueline A R Shea$^{\authref{chem-uc}}$,
Hyeondeok Shin$^{\authref{csd-anl}}$,
Luke Shulenburger$^{\authref{hedp-snl}}$,
Andreas Tillack$^{\authref{nccs-ornl}}$,
Joshua P Townsend$^{\authref{hedp-snl}}$,
Norm M. Tubman$^{\authref{chem-uc}}$,
Brett Van Der Goetz$^{\authref{chem-uc}}$,
Jordan E Vincent$^{\authref{phy-ui}}$,
D. ChangMo Yang$^{\authref{unist-korea}}$,
Yubo Yang$^{\authref{phy-ui}}$
Shuai Zhang$^{\authref{llnl}}$,
Luning Zhao$^{\authref{chem-uc}}$
}

\address{$^{\authlabel{intel}}$ Intel Corporation, Hillsboro, OR 987124, United States}
\address{$^{\authlabel{hedp-snl}}$ HEDP Theory Department, Sandia National Laboratories, Albuquerque, NM 87185, United States}
\address{$^{\authlabel{eng-uva}}$ Department of Mechanical and Aerospace Engineering, University of Virginia, Charlottesville, VA 22904}
\address{$^{\authlabel{alcf-anl}}$ Argonne Leadership Computing Facility, Argonne National Laboratory, Argonne, IL 60439, United States}
\address{$^{\authlabel{csd-anl}}$ Computational Science Division, Argonne National Laboratory, Argonne, IL 60439, United States}
\address{$^{\authlabel{phy-ncsu}}$ Department of Physics, North Carolina State University, Raleigh, NC 27695, United States}
\address{$^{\authlabel{csmd-ornl}}$ Computer Science and Mathematics Division, Oak Ridge National Laboratory, Oak Ridge, TN 37831, United States}
\address{$^{\authlabel{chem-cam}}$ University Chemical Laboratory,
  Lensfield Road, Cambridge, CB2 1EW, United Kingdom}
\address{$^{\authlabel{llnl}}$ Lawrence Livermore National Laboratory, Livermore,  CA 94550, United States}
\address{$^{\authlabel{sorbonne-fr}}$ Institut de Min\'eralogie, de Physique des Mat\'eriaux et de Cosmochimie (IMPMC), Sorbonne Universit\'e, CNRS UMR 7590, IRD UMR 206, MNHN, 4 Place Jussieu, 75252 Paris, France}
\address{$^{\authlabel{phy-ui}}$ Department of Physics, University of Illinois, Urbana, IL 61801, United States}
\address{$^{\authlabel{mrl-uc}}$ Materials Research Laboratory, University of California, Santa Barbara, CA, 93106, United States}
\address{$^{\authlabel{stonert}}$ Stone Ridge Technology, Bel Air, MD 21015, United States}
\address{$^{\authlabel{chem-brown}}$ Department of Chemistry, Brown University, Providence, RI 02912, United States}
\address{$^{\authlabel{msd-anl}}$ Material Science Division, Argonne National Laboratory, Argonne, IL 60439, United States}
\address{$^{\authlabel{ise-nw}}$ Northwestern-Argonne Institute for Science and Engineering, Northwestern University, Evanston, IL 60208, United States}
\address{$^{\authlabel{cnms-ornl}}$ Center for Nanophase Materials Sciences, Oak Ridge National Laboratory, Oak Ridge, TN 37831, United States}
\address{$^{\authlabel{csed-ornl}}$ Computational Sciences and Engineering Division, Oak Ridge National Laboratory, Oak Ridge, TN 37831, United States}
\address{$^{\authlabel{mstd-ornl}}$ Materials Science and Technology Division, Oak Ridge National Laboratory, Oak Ridge, TN 37831, United States}
\address{$^{\authlabel{nccs-ornl}}$ National Center for Computational Sciences, Oak Ridge National Laboratory, Oak Ridge, TN 37831, United States}
\address{$^{\authlabel{chem-uc}}$ Department of Chemistry, University of California, Berkeley, CA 94720, United States}
\address{$^{\authlabel{csd-lbnl}}$ Chemical Sciences Division, Lawrence Berkeley National Laboratory, Berkeley, California 94720, United States}
\address{$^{\authlabel{uw-parkside}}$ Department of Mathematics and Physics, University of Wisconsin-Parkside, Kenosha, WI 53144, United States}
\address{$^{\authlabel{unist-korea}}$ Center for Superfunctional
  Materials, Ulsan National Institute of Science and Technology, Ulsan
  44919, Republic of Korea}

\ead{kentpr@ornl.gov}
\begin{abstract}
  \qmcpack is an open source quantum Monte Carlo package
  for \textit{ab-initio} electronic structure calculations.
  It supports calculations of metallic and insulating solids,
  molecules, atoms, and some model Hamiltonians.  Implemented real space quantum
  Monte Carlo algorithms include variational, diffusion, and reptation Monte Carlo.
  \qmcpack uses Slater-Jastrow type trial wavefunctions in
  conjunction with a sophisticated optimizer capable of optimizing tens of thousands of parameters.
  The orbital space auxiliary field quantum Monte Carlo method is also implemented, enabling cross validation
  between different highly accurate methods. The code is
  specifically optimized for calculations with large numbers of
  electrons on the latest high performance computing architectures,
  including multicore central processing unit (CPU) and graphical processing
  unit (GPU) systems. We detail the program's
  capabilities, outline its structure, and give examples of its use
  in current research calculations. The package is available at
  \url{http://www.qmcpack.org}.
\end{abstract}

\vspace{2pc}



\section{Introduction}
\footnote{This manuscript has been authored by UT-Battelle, LLC under
  Contract No. DE-AC05- 00OR22725 with the U.S. Department of
  Energy. The United States Government retains and the publisher, by
  accepting the article for publication, acknowledges that the United
  States Government retains a non-exclusive, paid-up, irrevocable,
  world-wide license to publish or reproduce the published form of
  this manuscript, or allow others to do so, for United States
  Government purposes. The Department of Energy will provide public
  access to these results of federally sponsored research in
  accordance with the DOE Public Access Plan
  (\url{http://energy.gov/downloads/doe-public-access-plan}).}
An accurate solution of the many-body Schr\"odinger equation is a grand
challenge for physics and chemistry.  The great difficulty of
obtaining accurate yet tractable solutions has led to the development
of many complementary methods, each bearing unique approximations,
limitations, and assumptions. Today, the electronic structure of
periodic condensed matter systems is most commonly obtained using
density functional theory (DFT), while for isolated molecules
many-body quantum chemical techniques can also be
applied.\cite{martin_electronic_2004,martin_interacting_2016} With
these techniques, obtaining systematically improvable and increasingly
accurate results for general systems is a major challenge.  With DFT,
the challenge lies in deriving accurate approximations to and
constraints on the exact density functional, such as the recent
approximate SCAN functional\cite{sun_strongly_2015}.  For this reason,
a systematically improvable DFT is a challenging theory and therefore
progress is slow. In quantum chemistry, the most accurate methods are
systematically improvable but scale poorly with system size. They are
not well developed for periodic systems with hundreds of electrons,
and, in particular, are not yet suitable for describing metallic
states.  Other many-body methods, such as GW and dynamical
mean-field theory (DMFT), are limited by their approximations. GW is systematically
affected by a perturbative treatment of the electron-electron
interaction. DMFT, mostly used for ``correlated'' electronic
systems, suffers from the
local nature of its self energy despite being non-pertubative,
particularly when applied to low-dimensional
systems and/or to systems where a strong Hubbard repulsion is not the
only relevant contribution in the interaction.

Quantum Monte Carlo (QMC) methods provide an alternative
route to solutions of the many-body Schr\"odinger equation via
stochastic sampling\cite{BeccaSorellaBook2017}. By sampling the many-body wavefunction or its
projection, QMC methods largely avoid the need to perform numerical
integrals that scale poorly with system size. Further, QMC methods and
implementations generally invoke controllable approximations. Although
computationally expensive compared to DFT, QMC methods can be systematically improved and give nearly
exact results in some cases. This was most notably performed for the
homogeneous electron gas in 1980\cite{ceperley1980}. Besides their
stochasticity, the key distinctions between QMC and most other
electronic structure methods is that (1) for QMC the approximations
are few and well specified, and (2) QMC usually requires a ``trial
wavefunction'' as input.  The trial wavefunction is typically
constructed from the results of less costly methods, e.g. DFT or
small quantum chemical calculation, and
improved via subsequent optimization.  QMC methods can be directly
applied to materials and chemical problems of interest as with any
other electronic structure method but also serve as an important
validation tool to assess and improve the approximations of less
costly methods by providing reference benchmark data.

QMC methods have been applied to isolated molecules as well as
insulating, semiconducting, and metallic phases of condensed
matter. Complex
molecules\cite{ValssonJCTC2010,ZenPRB2016,CocciaJCTC2017},
liquids\cite{morales_quantum_2014}, molecular
solids\cite{mcminis_molecular_2015,ZenPNAS2018},
solids\cite{shulenburger_quantum_2013} and defect properties of
materials\cite{Parker2011,Hood2012,Ertekin2013,Santana2015,Santana2017}
have been studied at clamped nuclear geometries. Molecular dynamics
calculations driven by QMC nuclear forces, and calculations beyond the
Born-Oppenheimer approximation are also possible, e.g.,
\cite{tubman-beyond-born-2014,yang-beyond-born-2015,tubman-beyond-born-2016}. The
majority of results have been obtained with approaches that operate in
real space, using variational Monte Carlo (VMC) and diffusion Monte
Carlo
(DMC)\cite{foulkes_quantum_2001,dubecky_noncovalent_2016,wagner_discovering_2016,benali_development_2017,rubenstein_introduction_2017,saritas_investigation_2017}.
Whereas QMC methods that sample the many-body wavefunction in real
space have been in use for decades, there are an increasing number of
attractive methods that can be implemented in a basis of atomic
orbitals, such as auxiliary field
(AFQMC)\cite{ZhangPRB1997,pavarini_emergent_2013,Motta2017}, Monte
Carlo configuration interaction
(MCCI)~\cite{greer-montecarlo-1995,coe-montecarlo-2012} and
full-configuration interaction QMC (FCIQMC)\cite{booth_fermion_2009}.
Both real-space and basis approaches have different strengths and
weaknesses.  For example, more accurate multiple-projector
pseudopotentials and frozen core approaches are readily implemented in
AFQMC\cite{ma_auxiliary-field_2017,purwanto_frozen-orbital_2013} and
FCIQMC, but these methods are generally thought to have a higher
computational cost compared to real-space QMC.  Most significantly,
the increasing diversity of QMC methods with different approximations
will enable cross-validation of electronic structure schemes for
challenging chemical, physical and materials problems, and help guide
improvements in the methodology.

\qmcpack implements a variety of real-space solvers and a
complementary, recently developed AFQMC solver. The package is open
source and openly developed. \qmcpack is implemented in modern C++,
making strong use of object orientation and template based generic
programming techniques to facilitate high modularity, a separation of
functionalities, and significant code reuse. A special emphasis has
been given to performance, capability, and stability for large
production calculations. A state-of-the-art wavefunction optimization
algorithm capable of optimizing tens of thousands of parameters
enables the most accurate and sophisticated wavefunctions to be
utilized.\cite{zhao_blocked_2017} The latest size-consistent
algorithms for pseudopotential evaluation\cite{Casula2010} and time
step\cite{ZenPRB2016,ZenPNAS2018} are implemented.
The code is highly optimized
for modern high-performance computer architectures via extensive
vectorization, careful consideration of memory layouts and access
patterns\cite{mathuriya_optimization_2017}, efficient OpenMP
threading, and an implementation for graphics processing units (GPUs) using
NVIDIA's CUDA. A significant effort is underway to improve the code
for exascale architectures with a single common code base. QMC methods
are particularly attractive for these future systems due to the
relatively low data movement required. This combination of
capabilities and activities helps to distinguish \qmcpack from other
QMC codes such as QWALK\cite{wagner_qwalk_2009},
CASINO\cite{needs2010}, CHAMP\cite{umrigar_cornellholland_nodate},
and TurboRVB\cite{turborvburl}.

In this article, we give an overview of the features and capabilities
of the \qmcpack package.  To indicate the future development pathways,
we outline a number of challenges for QMC methods, including
development of consistent many-body pseudopotentials, the addition of
spin-orbit interactions to QMC Hamiltonians, and the challenge of
exascale computing.


\section{Open source and open development}

\qmcpack is open source and distributed under the Open Source
Initiative\cite{opensourceinitiativeurl} approved University of
Illinois/NCSA Open Source license. The main project website
\url{http://qmcpack.org} links to versioned releases and the
development source code. This also includes a substantial manual
detailing installation instructions, examples for
workstation through to supercomputer installations, and detailed methodology
and input parameter descriptions.  The source code includes a
substantial test framework ($> 300$ tests) including unit and
integration tests that are used to help validate the implementation and
test new installations.

\qmcpack is also openly developed. The latest source code and updates
are coordinated via GitHub,
\url{https://github.com/QMCPACK/qmcpack}. This site provides
version controlled source code, a wiki describing development
practices, issue (ticket) tracking, and a contribution review
framework (pull request reviews). The project follows the
``git flow'' branching and development model. 

Contributions from new developers are encouraged and follow exactly
the same mechanism as for established developers. For example, the ``finite
difference linear response'' (FDLR) method\cite{BluntJCP2017} was
recently contributed and underwent several updates to maximize
compatibility with the existing source code.  The full discussion
history of contributions is also available. The open development
process, with full change history available, allows contributions to
be clearly identified and credit accurately assigned. The source code
change history can be tracked to the earliest days of \qmcpack.

All proposed changes to \qmcpack automatically undergo continuous
integration testing which allows the contributions to be run on
different architectures and with different software versions than
might have been used for development, e.g. different processor
manufacturers, GPUs, or compilers. This allows for rapid feedback and
reduces risk that significant bugs are introduced.

Development directions are set in part based on requests from users
and experience applying \qmcpack to tutorial through to research level
problems. Besides contacting the developers directly or using GitHub,
a discussion group (``\qmcpack Google group'') provides a method to
make suggestions or obtain support. For example, requests to interface
\qmcpack with additional electronic structure or quantum chemistry
packages that will enable new science applications or solve an existing problem will be
given priority.


\section{Code structure}
\label{sec:structure}

\qmcpack is architected in a modular and generic structure, aiming to
facilitate the maximum reuse of source code and to appropriately abstract
key organizational and functional concepts. For example, as detailed
below, all the different components of a trial wavefunction utilize a
common base class and provide an identical interface to the different
QMC methods. This allows any newly contributed wavefunction component to become
immediately available to all the QMC methods.

Extensive use is made of C++ generic and template programming to
minimize reimplementation of common functionality. For example, the
numerical precision is parameterized, with single, double and mixed
precision implementation available largely from a single source
definition, and widely used functionality such as 1 to 3 dimensional
splines are accessed through a common interface.

In the following we give a high-level outline of the major
abstractions and components in the application. In practice and due to
the large range of functionality implemented, \qmcpack
consists of over 200 classes. To aid developers, the manual (\url{http://docs.qmcpack.org}) provides additional guidance while the Doxygen tool is
used to produce documentation to help track functionality and
interdependencies. Currently this is automatically generated from the
latest development source and published at
\url{http://docs.qmcpack.org/doxygen/doxy/}. 

\begin{figure}
\begin{center}
\includegraphics[width=0.7\columnwidth]{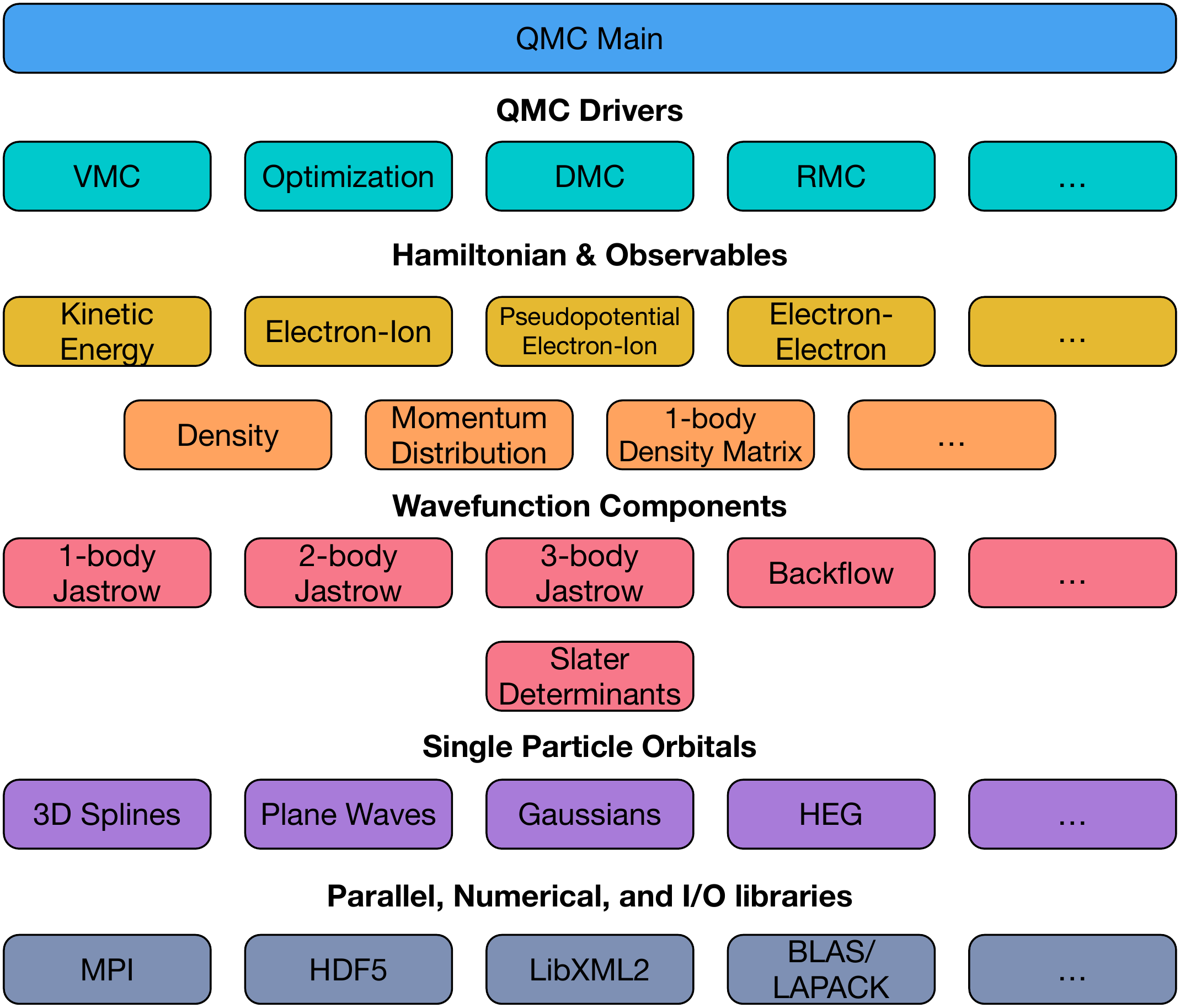}
\end{center}
\caption{High-level overview of the structure of \qmcpack. For simplicity many smaller components are not shown. This
  includes the particle classes, distance tables, and the branching and load balancing
  classes. The ``\ldots'' indicate additional high-level functionality
  is available. Dependencies between components flow from top to
  bottom, except for the libraries which are used by all component of
  the application.
\label{fig:qmcpack_structure}
}
\end{figure}

At a high level, \qmcpack consists of the following major abstractions
and areas of functionality (figure \ref{fig:qmcpack_structure}):
\begin{enumerate}
\item QMCMain. The topmost level of the application is responsible for
  parallel setup and initially parsing the input XML. Each XML section
  is handed to the appropriate functionality to setup the Hamiltonian
  or run a QMC calculation. Notably, the input driver persists walkers
  from section to section. A single \qmcpack input file can therefore
  describe a single simple VMC run, or considerably more complex
  and powerful workflows invoking VMC, wavefunction optimization, and production
  DMC calculations at a range of time steps. The user can choose the
  appropriate modality for their research.

\item QMC Drivers. These implement major QMC methods such as VMC,
  wavefunction optimization, reptation. Orbital space methods such as AFQMC
  (section \ref{sec:orbital_methods}) are also implemented here. Due to
  the levels of abstraction, the drivers have no dependencies on the
  specifics of the trial wavefunction that are in use.

\item Walkers and particles. Classes handle the state
  information for each walker, including the lists of particles that
  are updated by the Monte Carlo. Common infrastructure for computing
  minimum images in periodic boundary conditions is provided
  here. Walkers carry additional state information depending on the enabled
  Hamiltonian and Observables so that appropriate statistics can be
  reported.

\item Hamiltonians. The Hamiltonian used within in \qmcpack is described
  by the input XML. This enables model system calculations as well as first
  principles calculations to be performed. Beyond Born-Oppenheimer
  simulations are also supported, with the ionic positions a
  non-constant part of the Hamiltonian. The kinetic, bare
  electron-ion, pseudopotential, electron-electron and ion-ion 
  Coulomb terms are implemented at this level.

\item Observables. Quantities that are not critical to the evaluation
  of the Hamiltonian are termed observables and potentially include
  the density, density matrices, momentum distribution.

\item Wavefunction. The trial wavefunctions are implemented as a
   product of different wavefunction components. This includes single
   and multiple Slater determinants, and one,
   two, and three-body Jastrow terms. Specialized wavefunctions such
   as backflow and antisymmetrized geminal product (AGP) wavefunctions
   are also implemented here.

\item Single particle orbitals. Orbitals from plane-wave, spline, and gaussian basis
  sets are evaluated for use in e.g. Slater determinant components of
  the wavefunction. Specialty basis sets are also implemented,
  e.g. the plane-wave based homogeneous electron gas, and the hybrid
  augmented-plane wave basis set combining interstitial plane waves
  and atomic-core centred spherical harmonic expansions.

\item Standard libraries. Where available, standardized implementation
  and libraries are used for parallelization
  (MPI), I/O (HDF5, libxml2), linear algebra (BLAS/LAPACK), and
  Fourier transforms  (FFTW).

\end{enumerate}

\section{Performance and Parallel Scaling}
\label{sec:performance}
Due to the high computational cost of QMC methods, the \qmcpack
implementations have been heavily optimized to obtain a high on-node
performance and a high distributed parallel efficiency. Nevertheless, obtaining a
highly performing and efficient simulation remains an important
responsibility of the user, because a considered choice of QMC
methods, algorithms, accurate trial wavefunctions, and overall
statistics can significantly reduce the computational cost.

Many electronic structure methods obtain high computational efficiency
-- a high fraction of theoretical floating point performance -- via
use of dense linear algebra such as matrix multiplication. Real space
QMC methods are noteworthy for the relative lack of dense linear
algebra and a focus on particle-like operations, such as computing
inter-particle distances or evaluating small polynomial functions of particle
position. In this regard parts of QMC are similar to a classical
molecular dynamics code.

\begin{figure}
\begin{center}
  \includegraphics[width=0.7\columnwidth]{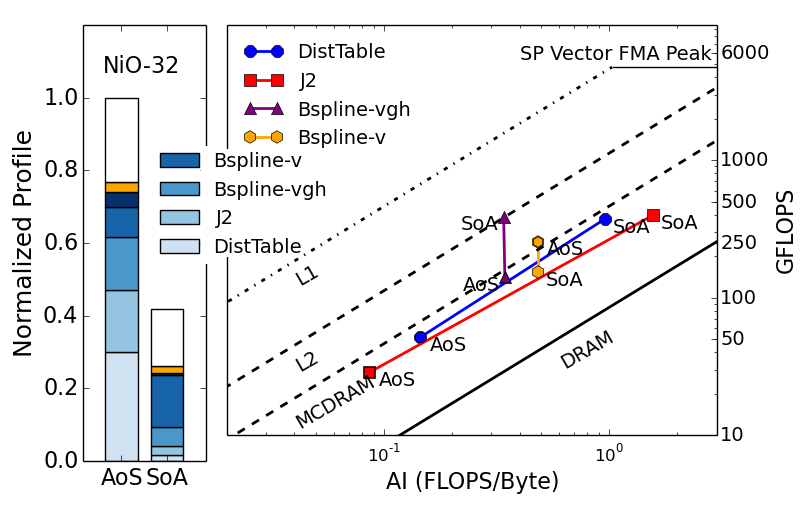}
\end{center}  
  \caption{Computational performance of key kernels in \qmcpack for an
    NiO 32 atom cell on Intel Knights Landing processors. By using a
    Structure of Arrays (SoA) layout and improving the implemented
    algorithms, higher arithmetic intensity (AI) is obtained compared
    to the Array of Structures (AoS) data layout used exclusively in
    older versions of QMCPACK. A significantly higher overall
    performance, measured in GFLOPS, is obtained in the new
    implementation.}
  \label{fig:roofline}
\end{figure}

To obtain high on-node performance, \qmcpack's implementations are
optimized to vectorize efficiently and to make efficient use of modern
memory hierarchies and maximize in-cache data reuse. For example,
while historically QMC codes have tended to avoid recomputing values,
for some operations it is now faster to compute properties on the
fly. This also reduces the memory requirements of the application. We
have recently completed extensive analysis and reimplementation of the
core compute kernels of the application, more than doubling the speed
of many calculations on modern multicore
processors\cite{mathuriya_optimization_2017}. The
performance obtained for several key kernels is shown in figure \ref{fig:roofline}.
To improve the computational efficiency of the largest calculations
with thousands of electrons where the Slater determinant update cost
is significant, we have recently proposed a delayed update algorithm
that enables increased use of matrix-matrix
multiplication\cite{McDanielJCP2017}.

\begin{figure}
  \includegraphics[width=0.49\columnwidth]{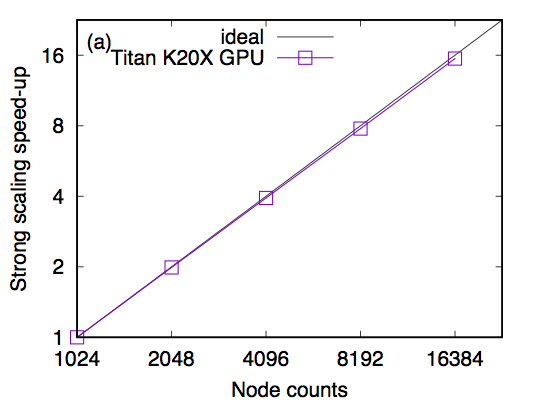}\quad
  \includegraphics[width=0.49\columnwidth]{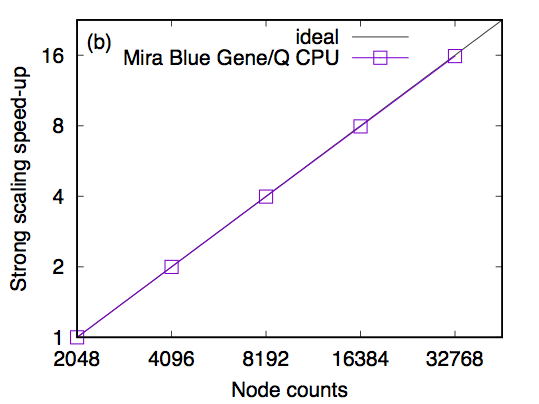}
  \caption{Parallel scaling of \qmcpack on two architectures for
    DMC calculations of a NiO 128 atom cell with 1536 valence
    electrons. Titan nodes have a single GPU each, and these runs used
    512000 total walkers. Each Blue
    Gene Q node has a 16 core processor, and these runs used 458752
    total walkers.}
  \label{fig:parallel_scaling}
\end{figure}

High parallel scalability is determined by exploiting parallelization
at two levels. First, on node parallelization is achieved through
OpenMP threading, or CUDA on GPUs. This allows common read-only data
such as trial wavefunction coefficients to be shared between threads,
reducing overall memory usage. Each OpenMP thread updates one or more
walkers, and multiple walkers can be assigned to each GPU. The second
level of parallelization is obtained via MPI. For simulations with a
variable number of QMC walkers, load balancing is performed by default
at every step. Asynchronous messaging is used to reduce the time to
load balance all the walkers across all the nodes, but in the current
implementation a global reduction is still required to compute the
ensemble average energy needed for load balancing. As shown in figure
\ref{fig:parallel_scaling}, even for modest calculations the
scalability is sufficient to fully utilize the largest
supercomputers. The first level of parallelization within each node
reduces the total volume of MPI messaging because the load balancing
only needs to be performed at the per-node level. Options to adjust
the load balancing frequency and alternative algorithms such as
stochastic reconfiguration\cite{AssarafPRE2000} are available.

\section{Real space QMC methods}
\label{sec:real_space_methods}

\subsection{Introduction}
The main production algorithms in \qmcpack use methods based on
Monte Carlo sampling of electron positions in real space to produce
highly accurate estimates of the many-body ground-state wavefunction
$|\Phi_0\rangle$ and its associated properties. Note that QMC also has
complementary \textit{orbital space}-based approaches that work within
second quantization, as described in section \ref{sec:orbital_methods}.

VMC and DMC are the most commonly applied real-space QMC methods. Within VMC, the simplest scheme, Monte Carlo sampling is used to
obtain estimates of the energy of a trial wavefunction
\begin{equation}\label{eqn:vmc}
E_{VMC}=\int \Psi^* \hat{H} \Psi  \;.
\end{equation}
In DMC, the ground-state wavefunction is obtained by projection of the imaginary time Schr\"{o}dinger equation
\begin{equation}\label{imaginary_time_schrodinger}
	-\frac{\partial \Psi}{\partial \beta} = \hat{H}\Psi
\end{equation}
to long time,  where $ \beta = it $ and has units of imaginary
time. (Hartree units are used here and throughout, except where noted.) Crucial to both
methods is an accurate trial or guiding wavefunction. Clearly, in VMC the
trial
wavefunction completely determines the accuracy and statistical efficiency
of the result. In DMC it is the nodal surface of the trial
wavefunction that determines the accuracy, while the overall trial
wavefunction determines the statistical efficiency. The full range of
supported trial wavefunctions is described in
section\;\ref{sec:trialwavefunctions}.

The real-space methods use a Hamiltonian within the Born-Oppenheimer approximation (BOA):
\begin{equation}\label{vmc:electron_ham}
	\hat{H}  = -\frac{1}{2} \sum_i \nabla^2_i + \frac{1}{2} \sum_{i\neq j} \frac{1}{|{\mathbf{r}_i} - {\mathbf{r}_j}|} + \sum_{i,J} v^{eJ}({\mathbf{r}_i},\mathbf{R}_J)
\end{equation}
where the lower case indices and positions ${\mathbf{r}}_i$ refer to the electrons, and the upper case indices and positions $\mathbf{R}_I$ refer to the ions.
In order, the terms in \ref{vmc:electron_ham} correspond to the
kinetic energy of the electrons, the potential energy of the
electrons, and the potential energy due to interactions between
electrons and ions. The energy contribution due to the Coulomb
interactions of the atoms is constant within the BOA, and is computed
by an Ewald sum. Further details are given in section \ref{sec:hamiltonian}

For technical reasons, in the following we will work with the ``importance sampled" Schr\"{o}dinger equation, which can be obtained from the imaginary time Schr\"{o}dinger equation by rewriting it in terms of $f(\mathbf{r},\beta) = \Psi_T(\mathbf{r})\Psi(\mathbf{r},\beta)$.   


\begin{eqnarray} \label{pmc:impsamp_schrodinger}
\frac{\partial f}{\partial \tau}&  =& \hat{L} f(\mathbf{r},\beta) \\
                                   &=& \left[ \lambda_e \nabla \cdot (\nabla - \mathbf{F}(\mathbf{r})) - (E_L(\mathbf{r})-E_T) \right]f(\mathbf{r},\beta)
\end{eqnarray}

where $\Psi_T(\mathbf{r})$ is the trial wavefunction,
$\mathbf{F}(\mathbf{r})=2\nabla \log \Psi_T(\mathbf{r})$ is the
``wavefunction force", and
$E_T(\mathbf{r})=\Psi_T(\mathbf{r})^{-1}\hat{H}\Psi_T(\mathbf{r})$ is
the ``local energy". $\hat{L}$ is the ``importance-sampled
Hamiltonian" operator. To help future discussion, we split $\hat{L}$
into a ``drift/diffusion" operator $\hat{K}$ and a ``branching"
operator $\hat{E}$, given by

\begin{equation} \hat{K} =
  \frac{1}{2}\nabla \cdot \left(\nabla -
    \mathbf{F}(\mathbf{r})\right)
\end{equation}
\begin{equation}
  \hat{E} = - \left(E_L(\mathbf{r}) - E_T\right)
\end{equation}

One advantage of working with the physical Hamiltonian as opposed to an auxiliary problem (as in Kohn-Sham DFT), is that the variational theorem of quantum mechanics holds, which states that for a trial wavefunction $\Psi_T$,   
\begin{equation}\label{vmc:var_energy}
E_{\Psi_T} = \frac{\langle \Psi_T | \hat{H} | \Psi_T \rangle }{\langle \Psi_T | \Psi_T \rangle} = \frac{\langle \Psi_T | E_L(\hat{\mathbf{r}}) | \Psi_T \rangle }{\langle \Psi_T | \Psi_T \rangle} \geq E_0
\end{equation}
The strict equality holds if $\Psi_T$ is the ground-state of $\hat{H}$.  A corollary is that the variance of the trial wavefunction also obeys the following:
\begin{equation}\label{vmc:var_variance}
\sigma^2_{\Psi_T} = \frac{\langle \Psi_T | \hat{H}^2 - (E_{\Psi_T})^2 | \Psi_T \rangle}{\langle \Psi_T | \Psi_T \rangle} \geq 0
\end{equation}

The variational principle is significant, since it gives us a well-defined metric for which wavefunctions are better or worse approximations to the ground state.  This can be turned into an actual algorithm by parameterizing families of wavefunction ansatz with parameters $\mathbf{c}$.  One can then minimize equation \ref{vmc:var_energy} with respect to $\mathbf{c}$ to obtain a best estimate for the state.
 
 \subsection{Variational Monte Carlo}
The oldest approach for dealing with the Schr\"odinger equation 
for realistic systems involves writing  an approximation for the
ground-state wavefunction and evaluating expectation values.  There
are two ingredients in this procedure:  evaluating
equation \ref{vmc:var_energy} for some set of variational parameters
$\mathbf{c}$, and then minimizing.  The optimization procedure is
covered in detail in Section \ref{sec:optimization}. 

The energy expectation value in equation \ref{vmc:var_energy}, as well as
all physical expectation values, are integrals of a form that are
amenable to Metropolis Monte Carlo sampling.  Thus, we can evaluate
equation \ref{vmc:var_energy} (for example) by the following:

\begin{equation}
E_{\Psi_T} = \frac{\int d\mathbf{r} |\Psi_T(\mathbf{r})|^2 E_L(\mathbf{r})}{\int d\mathbf{r} |\Psi_T(\mathbf{r})|^2} = \frac{1}{N_s} \sum_{i} E_L(\mathbf{r}(t_i)) + \xi
\end{equation}

$N_s$ is the number of samples, and $\xi$ is a Gaussian-distributed
statistical error whose variance scales like $1/\sqrt{N_s}$.  We write
the sample configurations as $\mathbf{r}(t_i)$ to emphasize that
Metropolis Monte Carlo generates samples sequentially via a random
walk along a Markov chain. To parallelize the algorithm, multiple
independent Markov chains or ``walkers'' are used.

\subsubsection{Trial Moves}
\qmcpack supports VMC trial moves with and without drift.  This means that the move $\mathbf{r}\to\mathbf{r}'$ is drawn from the transition probability distribution given by:

\begin{equation}\label{trans_prob_alle}
T(\mathbf{r}\to\mathbf{r}',\tau) = \frac{1}{(4\pi\lambda\tau)^{3N/2}}\exp \left(-\frac{\left(\mathbf{r'} -\mathbf{r} - 2\lambda\tau \mathbf{F}(\mathbf{r})\right)^2}{4\lambda \tau}\right)
\end{equation}

For drift based moves, $\mathbf{F}(\mathbf{r})$ is taken to be the
same wavefunction force as appears in
equation \ref{pmc:impsamp_schrodinger}.  For moves without drift,
$\mathbf{F}(\mathbf{r})=0$.  In the absence of pathologies in the
trial wavefunction, the use of the drift term is almost always more efficient.  

In addition to drift or no-drift based moves, the code supports particle-by-particle or all-electron moves.  All-electron moves are conceptually the simplest.  One proposes the move $\mathbf{r}\to\mathbf{r}'=\mathbf{r}+\mathbf{\Delta}$ by drawing the $3N_e$ dimensional vector $\mathbf{\Delta}$ from the distribution in equation \ref{trans_prob_alle}.  This move is then accepted or rejected with probability:
\begin{equation}\label{accept_reject}
A(\mathbf{r}\to\mathbf{r}') = \min \left(1.0, \frac{|\Psi_T(\mathbf{r}')|^2}{|\Psi_T(\mathbf{r})|^2}\frac{T(\mathbf{r}'\to\mathbf{r})}{T(\mathbf{r}\to\mathbf{r}')} \right)
\end{equation}

In contrast, particle-by-particle moves work by iterating sequentially over all electrons.  Considering an electron $i$ at position $\mathbf{r}_i$.  A particle-by-particle move is executed by first drawing a new position for electron $i$ from the following probability distribution.
\begin{equation}
T(\mathbf{r}_0,\ldots,\mathbf{r}_i \to \mathbf{r}'_i,\ldots,\mathbf{r}_{N_e})=\frac{1}{(4\pi\lambda\tau)^{3/2}}\exp \left(-\frac{\left(\mathbf{r'}_i -\mathbf{r}_i - 2\lambda\tau \mathbf{F}_i(\mathbf{r})\right)^2}{4\lambda \tau}\right)
\end{equation}
Then, the move is accepted or rejected using a similar acceptance
probability as in equation \ref{accept_reject}.  Particle-by-particle moves
are typically favored over all-electron moves, due to their higher
statistical and numerical efficiencies in practice. However, all-electron moves may be competitive for small
systems or for sophisticated trial wavefunctions where single particle
moves can not be cheaply evaluated numerically.

\subsection{Projector Monte Carlo}
One can substantially improve upon the accuracy of VMC by using
projector Monte Carlo methods such as DMC.  The ``projector" is the formal solution of the imaginary time Schr\"{o}dinger equation $\hat{G}(\beta) = \exp(-\beta\hat{H})$, and has the very desirable property that given any trial wavefunction $|\Psi_T\rangle$ which is non-orthogonal to the ground-state wavefunction, one can obtain the ground-state $|\Phi_0\rangle$ by the following:
\begin{equation}
\lim_{\beta \to \infty} \hat{G}(\beta)|\Psi_T\rangle = e^{-\beta E_0}|\Phi_0\rangle
\end{equation} 

For efficiency reasons, we consider the projector $\tilde{G}(\mathbf{r},\mathbf{r}',\beta)$ associated with the importance sampled Schr\"{o}dinger equation.  For realistic systems, it is exceedingly rare to have exact analytic expressions for the projector. However, we can solve for the Green's function $\hat{\tilde{G}}(\tau)$ of equation \ref{pmc:impsamp_schrodinger} 
approximately for short times $\tau$. 
Solving the drift/diffusion equations and rate equations independently in the short-time limits, one uses the 
symmetric Trotter formula:
\begin{equation} 
\exp \left(\tau(\hat{A}+\hat{B})\right)=\exp \left(\frac{\tau}{2}\hat{B} \right)\exp \left(\tau\hat{A} \right)\exp \left(\frac{\tau}{2}\hat{B}\right)+O(\tau^2)
\end{equation}
to stitch these independent solutions together into an approximate solution for the importance sampled Green's function:

\begin{equation}\label{pmc:g_shorttime}
\tilde{G}(\mathbf{r},\mathbf{r}',\tau) = \bra{\mathbf{r}}\hat{\tilde{G}}(\tau)\ket{\mathbf{r}'} = G_{DD}(\mathbf{r},\mathbf{r}',\tau) G_{B}(\mathbf{r},\mathbf{r}',\tau)+O(\tau^2)
\end{equation} 

$G_{DD}(\mathbf{r},\mathbf{r}',\tau)$ is the Green's function for the drift/diffusion 
operator $\lambda_e \nabla \cdot (\nabla - \mathbf{F}(\mathbf{r}))$.  
Assuming that $\mathbf{F}(\mathbf{r})$ is slowly varying, its solution is given by:
\begin{equation}\label{pmc:g_driftdiffusion}
\tilde{G}_{DD}(\mathbf{r},\mathbf{r'},\tau) = \frac{1}{(4\pi\lambda\tau)^{3N/2}}\exp \left( - \frac{(\mathbf{r'}-\mathbf{r} - 2\lambda \tau \mathbf{F}(\mathbf{r}) ) ^2}{4\lambda \tau} \right)
\end{equation}

The Green's function for the local energy operator is:
\begin{equation}\label{pmc:g_growth}
\tilde{G}_B(\mathbf{r},\mathbf{r}',\tau) = P_0 \exp \left( - \frac{1}{2}(E_L(\mathbf{r})+E_L(\mathbf{r}') - 2E_T)\tau\right)
\end{equation}

Near the nodes of $\Psi_T(\mathbf{r})$ and near bare ions,
singularities render the ``slowly-varying" approximation used in
equation \ref{pmc:g_driftdiffusion} invalid.  Improved drift-diffusion
projectors have been derived which have been shown to reduce the time
step error \cite{umrigar1993}.  \qmcpack implements drift rescaling
based on proximity to the nodal surface, following the prescription in
\cite{umrigar1993} for single-electron and all-electron moves.
Rescaling based on proximity to bare ions is not yet implemented. 

\subsection{Diffusion Monte Carlo}

Diffusion Monte Carlo works by stochastically simulating the imaginary-time evolution of an initially prepared state 
$f(\mathbf{r},0)=|\Psi_T(\mathbf{r})|^2$.  This is done by exploiting the mathematical
correspondence between Fokker-Planck equations
for the evolution of probability distributions, and Langevin equations describing
the stochastic evolution of particle trajectories.  
To shift to a Langevin picture, we represent an initial state
$f(\mathbf{r},0)$ by an ensemble of $N_w$ walkers 
$\lbrace \mathbf{r}_0(0),\mathbf{r}_1(0),\ldots,\mathbf{r}_{N_w-1}(0) \rbrace$ distributed
according to $f(\mathbf{r},0)$.  Assume each walker also has an associated weight $w_i(\beta)$ where $w_i(0)=1.0$.  
We consider the action of the short-time Green's function $\tilde{G}(\mathbf{r},\mathbf{r}',\tau)$
on this distribution.

The action of the drift-diffusion propagator can be simulated with a random drift-diffusion step, given by:
\begin{equation}\label{pmc:driftdiffusion_step}
\mathbf{r}(\beta+\tau) = \mathbf{r}(\beta) + 2\lambda\tau\mathbf{F}(\mathbf{r}(\beta)) + \sqrt{2\lambda\tau} \bxi
\end{equation}

Here, $\bxi$
 is a $3N_e$ dimensional Gaussian random vector with unit variance.  Once a new position generated, 
the $G_B(\mathbf{r},\mathbf{r}',\tau)$ contribution is 
dealt with by updating the walker weight $w_i(\beta)$ with the following formula: 
\begin{equation}\label{pmc:walker_weight}
w(\beta+\tau) = w(\beta)G_B(\mathbf{r}(\beta),\mathbf{r}(\beta+\tau),\tau)
\end{equation}

Expectation values of local observables $\mathcal{A}(\mathbf{r})$ over the distribution $f(\mathbf{r},\beta)$ are obtained by 
a weighted average:
\begin{equation}\label{pmc:weighted_average}
\langle\mathcal{A}\rangle_f(\beta)=\frac{\sum_{i=0}^{N_w-1} w_i(\beta)\mathcal{A}(\mathbf{r_i(\beta))}}{\sum_{i=0}^{N_w-1} w_i(\beta)}
\end{equation}

Implementing everything discussed up to this point results in ``pure diffusion" Monte Carlo.  
However, due to the exponential growth/decay
of the walker weights with $\beta$, the efficiency of this method decays
exponentially with the projection time.  ``Branching" diffusion
Monte Carlo circumvents this problem by implementing equation \ref{pmc:walker_weight}
stochastically through the replication/removal of
walkers.  For each walker $i$, $M_i$ copies of the walker are made after the
drift/diffusion step according to the formula 
$M_i = \textrm{INT}(w_i(\beta+\tau)+\xi)$, where $\xi$ is a uniform random number
between 0 and 1.  The weights of these $M_i$ walkers are renormalized
and the copies then proceed to the next time step.  Notice that $M_i=0$ implies that the walker is killed.  

To avoid a walker population explosion or collapse, practical DMC
simulations adjust $E_T$ dynamically to keep the population finite and
stable. In \qmcpack this is achieved using either a variable number of
walkers combined with the above population control, or via a fixed
walker count scheme (``stochastic
reconfiguration''\cite{AssarafPRE2000}).
Both schemes can potentially introduce a ``population control bias" that must be checked and controlled
for, particularly for small populations. To minimize or check for
bias, the population should
be as large a possible, should be allowed to fluctuate
significantly, and the simulation run for a long time. 

To correctly simulate Fermionic systems and avoid a
collapse of the propagated wavefunction to a Bosonic solution, the ``fixed
node approximation'' is implemented.  This constrains the projected
solution to the nodal surface of the trial wavefunction, thereby
preserving Fermionicity. Proposed moves that result in a nodal crossing
are detected through the change in sign of the wavefunction and are
rejected. This is usually the most significant
approximation made, and requires accurate trial wavefunctions.

Finally, we note that several important modifications for production
calculations: \qmcpack implements the ``small time-step error''
algorithm due to Umrigar et al. \cite{umrigar1993}, where the drift
term is modified near wavefunction nodes and effective time step
introduced to improve the time step convergence of the algorithm. The
recently proposed size-consistent
variation\cite{ZenPRB2016,ZenPNAS2018} is also implemented. This is
particularly effective for computing energy differences between very
different sized-systems, such as absorption energies or the formation
energies of molecular crystals\cite{ZenPNAS2018}.

\subsection{Reptation Monte Carlo}
Reptation Monte Carlo is constructed by exploiting the Feynman
path-integral formulation of Schr\"{o}dinger's equation\cite{BaroniPRL1999}.  Its primary
advantages over DMC are its ability to estimate observables over pure
distributions in polynomial time, and lack of population bias and
control issues.   Consider the ground-state ``partition function'':
\begin{equation}
\mathcal{Z}(\beta) = \langle \Psi_T | e^{-\beta\hat{H}} | \Psi_T \rangle
\end{equation}
First, we split $e^{-\beta\hat{H}}$ into $n$ segments each spanning an imaginary time $\tau=\beta/n$. After inserting n+1 position space resolutions of the identity and rewriting the resulting expression in terms of the importance sampled projector, we find that $\mathcal{Z}(\beta)$ can be written as:
\begin{eqnarray}
\mathcal{Z}(\beta) & =& \int d\mathbf{r}(t_0)\ldots d\mathbf{r}(t_n) \mathcal{P}[X] e^{-\sum_{i=0}^{n-1} L_{DMC}(\mathbf{r}(t_i),\mathbf{r}(t_{i+1}) } \\
\mathcal{P}[X] &=& |\Psi_T(\mathbf{r}(t_0)|^2 \tilde{G}_{DD}(\mathbf{r}(t_0),\mathbf{r}(t_1)) \ldots \tilde{G}_{DD}(\mathbf{r}(t_{n-1}),\mathbf{r}(t_n)) \\
L_{DMC}(\mathbf{r},\mathbf{r}') &=& \frac{\tau}{2} \left( E_L(\mathbf{r}) + E_L(\mathbf{r}') -2E_T\right)
\end{eqnarray}

X is shorthand for a ``path"
$X=(\mathbf{r}(t_0),\ldots,\mathbf{r}(t_n))$.  The reader will
recognize $\mathcal{Z}(\beta)$ as a path integral.  $L_{DMC}$ is called the ``link action", which when summed over $t$, gives the action for the path $X$.  $\mathcal{P}[X]$ is the probability of a walker which is initially distributed according to $|\Psi_T(\mathbf{r})|^2$ executing the directed random walk from $\mathbf{r}(t_0)$ to $\mathbf{r}(t_n)$ along the path $X$.

All reptation moves in \qmcpack use the ``bounce algorithm''\cite{CeperleyBounce}. For move
proposals, the improved propagators described in the DMC section are
directly used in reptation. In addition to nodal drift-rescaling, we
incorporate the DMC effective time step and energy filtering methods
directly into the link-action, which helps to significantly reduce
ergodicity problems associated with reptiles getting stuck in low
energy regions of configuration space. In the event that reptiles
still get stuck, the age of all reptile beads is accumulated. If a
bead exceeds some specified age, the entire reptile is forced to
propagate for $n$ steps without rejection and then re-equilibrate.

Since $\mathcal{P}[X]$ cancels out of the reptile accept/reject step,
any all-electron move which is a valid VMC configuration is supported
in RMC. In addition to traditional all-electron moves, \qmcpack also
supports reptile proposals which are built from a sequence of $N_e$
particle-by-particle moves. Reptile moves proposed with these
``particle-by-particle'' moves exhibit higher acceptance ratios than
the traditional all-electron moves, and are thus favored if memory is
available.

Propagation in RMC is supported for all-electron, local, and
semi-local pseudopotential Hamiltonians. The fixed-node
constraint is enforced by rejecting proposed
node crossings and immediately bouncing.

Since the random walk of each reptile is totally independent of other
reptiles, RMC is straightforwardly parallelized. In addition to generic MPI
parallelization, \qmcpack's RMC driver is able to place one reptile per
OpenMP thread on shared memory systems.


\section{Trial Wavefunctions}
\label{sec:trialwavefunctions}

Within QMC methods, the goal of the trial wavefunction is to represent
the true Fermionic many-body wavefunction of the studied system as
accurately as possible, including all the correlated electron
physics. Due to the large number of evaluations of the wavefunction
values and derivatives during the Monte Carlo sampling, it is also important that
the trial wavefunction be computationally cheap enough and use little
enough memory in order to be practical. These
are different considerations from those applied in DFT and in
the more closely related quantum chemical methods, leading to
different preferences.

Several different trial wavefunction forms are implemented in \qmcpack,
with varying suitability for solid state and molecular systems, and
different trade-offs between accuracy, memory usage, and number of
parameters. The most common form is the multi-determinant
Slater-Jastrow form (section\;\ref{sec:mdsj}), where the orbitals in each
determinant are evaluated using either real space splines or a
Gaussian basis set (section\;\ref{sec:orbitals}). The orbitals are usually
obtained from a mean-field method and imported to \qmcpack. The determinantal part
ensures that the trial wavefunction is properly antisymmetric with
respect to exchange of electron positions, i.e. Fermionic. Additional
correlations are incorporated via a symmetric real-space Jastrow
factor (section\;\ref{sec:jastrow}).  The Jastrow factor is usually
obtained via optimization entirely within \qmcpack (section\;\ref{sec:optimization}).

\subsection{Multi-Determinant Slater-Jastrow Form}
\label{sec:mdsj}

For the vast majority of molecular and solid-state studies, the
trial wavefunction is written as the product of an antisymmetric
function and a symmetric Jastrow function
\begin{equation}
  \label{eq:msdj}
  \Psi_T=\sum_{i=1}^{M}c_{i}D^\uparrow_iD^\downarrow_i e^J,
\end{equation}
where the $N$ electron trial wavefunction $\Psi_T$ is expanded in a
weighted sum of products of up and down spin determinants, $D$. These
are in turn multiplied by a real-space Jastrow factor, $J$. When
exponentiated, this factor is nodeless and the nodes of the trial
wavefunction are therefore purely determined by the determinantal
parts. A single product of up-spin and down-spin determinants would
correspond to a mean-field or Hartree-Fock starting point. Larger
determinantal sums can be obtained, e.g., from multi-configuration
self-consistent field quantum chemical calculations, CIPSI (section
\ref{sec:cipsi}), or be constructed based on physical or chemical
reasoning. Excited states may be constructed by manipulating the
occupancy of the Slater determinants in the input, e.g. to create an
exciton. Wavefunctions with greater or fewer electrons than the
neutral ground state may be similarly prepared to compute electronic
affinities or ionization potentials.

Due to the potentially large computational cost in evaluating the
trial wavefunction, \qmcpack uses previously computed data and
optimized methods to avoid full recomputation wherever effective and
practical. For single electron moves, \qmcpack uses the
Sherman-Morrison algorithm, as described in \cite{FahyPRB1990}. For
large calculations with thousands of electrons, the delayed update
scheme of \cite{McDanielJCP2017} is currently being implemented. For
calculations with multiple determinants, \qmcpack implements the
``table method'' of Clark et al.\cite{ClarkJCP2011}. This exploits the
relationship between largely similar determinants to cheaply compute
the determinant values while only requiring the full $N^2$ memory cost
of a single determinant. This enables, e.g., molecular calculations
that approach or even reach ``chemical accuracy'' to be
performed\cite{morales_multideterminant_2012}. To date, calculations
with up to $O(10^6)$ determinants have been performed (see
section\;\ref{sec:cipsi}), with larger calculations clearly
possible\cite{assaraf_optimizing_2017}.

\subsection{Orbitals}
\label{sec:orbitals}

The single-particle orbitals in the Slater determinants are generally determined by another
electronic structure code and imported into \qmcpack for calculations.  
\qmcpack has an easily extensible mechanism for adding new ways of
representing single-particle orbitals.  This can be particularly
useful when addressing model systems or performing specialized
tests. For example, \qmcpack supports specialty homogeneous electron
gas and plane-wave based wavefunctions, and for work on spherical quantum dots,
radial numerical functions\cite{BajdichPRB2011}. However, by far the
most common sources of orbitals are plane-wave based from Quantum
Espresso (QE) \cite{QuantumEspresso2009,QuantumEspresso2017},  and Gaussian based from the GAMESS
code\cite{gordon05}. Converters from these codes are provided and can
straightforwardly be extended to other methodologically related codes.

\subsubsection{B-spline basis sets}
\label{sec:bspline}
For calculations involving periodic boundary conditions, the standard
route is to first perform a DFT calculation using QE and
to then import the plane-wave coefficients into \qmcpack. Finite
molecular systems can also be studied by adding a considerable vacuum
region. \qmcpack then allows the boundary conditions to be made
aperiodic, even for orbitals originally based on plane-waves.

Although the single-particle orbitals can be evaluated directly in
the plane-wave basis, this requires evaluating each plane-wave for
every orbital and is thus very expensive: the cost grows with the
number of plane-waves.  For this reason, the single-particle orbitals
are usually converted into a regular 3D B-spline representation in
real-space. As implemented, this requires a constant 64 coefficients
to be accessed in memory to evaluate each single-particle wavefunction
regardless of the size of the underlying basis. These
operations are optimized to vectorize very well on current computer architectures,
enabling the orbital evaluation to run very efficiently. 

The principal downside of a B-spline basis is memory consumption,
particularly for large simulation cells.  Naively, the memory cost scales
as O($N^2$). For larger calculations the B-spline tables can
easily grow to tens or even hundreds of gigabytes, potentially
exceeding available memory.  Currently \qmcpack
shares the B-spline table among all processors on a node (or
GPU), but memory limitations can still constrain the calculations that can
be performed.  In the case of supercell calculations, \qmcpack can
exploit Bloch's theorem to reduce the demand. To save additional
memory, the spline coefficients may also be stored in single
precision, halving the amount of memory required compared to the full
double precision used in the originating plane-wave code. However,
memory usage of B-splines remains a problem for large simulation cells.

To further reduce memory costs,
\qmcpack can utilize a hybrid basis set composed of radial
splines times spherical harmonics near the atoms and B-splines elsewhere
in space.\cite{EslerPRL2010,EslerCISE2012} This is similar
to the augmented plane-wave schemes used by some DFT
implementations. The scheme allows for the high frequency components of the trial
wavefunction near the atomic nuclei to be
represented by a compact radial function and the smoother part of the
wavefunction in the interstitial regions to be represented by a much
coarser B-spline table.  The hybrid basis can save a factor of 4-8 in memory compared to the
standard B-spline representation while maintaining accuracy. Obtaining the hybrid
representation from a plane-wave basis requires an initial computationally costly conversion.

\subsubsection{Gaussian basis sets}

For molecular systems, one typically uses a Gaussian basis set to
represent the single-particle orbitals. \qmcpack supports standard
quantum chemical basis sets including contractions and for arbitrary
angular momenta. Atomic or natural orbitals can therefore be directly
imported from standard quantum chemistry codes. Interfaces currently
exist to GAMESS\cite{gordon05}, quantum
package\cite{QP}, and for packages supporting the
MOLDEN format. Interfacing requires converting the output of the
intended package to \qmcpack's XML or HDF5 format. For all-electron
calculations, a cusp correction scheme is implemented to enforce the
electron-nuclear cusp.

\subsubsection{Specialized basis sets}

Besides the B-spline and Gaussian basis sets described above, \qmcpack
implements several additional specialized basis sets for specific
problems. This includes Slater Trial Orbitals (STOs), the homogeneous
electron gas, and radial numerical functions for atomic
calculations. Due to the flexible internal architecture, orbitals can
be expressed in any combination of these functions. For example,  in \cite{KrogelJCP2018}, it was proposed to save memory by
storing orbitals on different sets of B-spline tables based on their
kinetic energy. This scheme did not require any source code
modifications.

\subsection{Backflow wavefunctions}

Improvement of the nodal surface can be achieved through backflow
wavefunctions, complementing the multideterminant route.
The formal justification for backflow wavefunctions rests on the
homogeneous electron gas and Fermi liquid theory \cite{Kwon1993}. Backflow
appears promising for bulk applications\cite{RiosPRE2006},
and has also been shown to aid in capturing dynamical
correlations in molecular systems when used in conjunction with
multideterminant wavefunctions \cite{Seth2011}.

Backflow wavefunctions are constructed from determinantal
wavefunctions as follows.  Instead of evaluating the Slater matrix
$M_{ij}=\phi_j (\mathbf{r}_i)$ at the bare electron coordinates
$\mathbf{r}_i$, we evaluate it at new quasiparticle coordinates 
$\tilde{M}_{ij}=\phi_j (\mathbf{q}_i)$.  The ``backflow
transformation'' from $\mathbf{r}_i\to \mathbf{q}_i$ is defined as:
\begin{equation}
\mathbf{q}_{i_\alpha}=\mathbf{r}_{i_\alpha}+\sum_{\alpha \leq \beta} \sum_{i_\alpha \neq j_\beta} \eta^{\alpha\beta}(|\mathbf{r}_{i_\alpha}-\mathbf{r}_{j_\beta}|)(\mathbf{r}_{i_\alpha}-\mathbf{r}_{j_\beta})
\end{equation}

In \qmcpack, the $\eta^{\alpha\beta}(r)$ are short-ranged, spherically
symmetric functions represented by fully optimizable
B-splines. \qmcpack allows for separate optimization of same-spin,
opposite-spin, and electron-ion terms.  Currently, backflow is fully
supported only with single determinant wavefunctions, but it can be
used in both bulk and molecular systems.


\section{Jastrow factors}
\label{sec:jastrow}

Jastrow factors\cite{jastrow1955}  are included in the trial wavefunction to improve the
representation of the many-body wavefunction. This non-negative Bosonic factor is
in principle an arbitrary function of all electron and ionic
positions, but in practical calculations are most commonly built from
functions systematically incorporating one, two, and three-body
correlations.  Notably, the Jastrow factor can readily satisfy the
electron-electron and electron-nucleus cusp conditions
\cite{kato1957,pack1966}, which are very slow to converge in the
multideterminant expansions commonly used in quantum chemistry.  The
improved representation of the many-body wavefunction naturally
reduces the statistical variance of the local energy and also
improves the quality of the DMC projection
operator\cite{grimm1971,kalos1974}, which is useful in the context of
timestep and nonlocal pseudopotential localization errors.

The bosonic ground state for $N$ particles can be written
\begin{eqnarray}
  \Psi_B = e^{-J(\mathbf{R})}
\end{eqnarray} 
with $J$ symmetric and where $\mathbf{R}$ denotes all the particle positions.  For fermions, the fixed node \cite{anderson1975,anderson1976} (or fixed phase \cite{ortiz-fp93}) wavefunction that arises from DMC projection has a related form.  In this case, a Jastrow wavefunction appears as a prefactor\cite{holzmann2016} modifying the local structure of the input Fermionic trial wavefunction, $\Phi_T$ to account for many-body correlations:
\begin{eqnarray}
  \Psi_{FN} = e^{-J(\mathbf{R})}\Phi_T(\mathbf{R})
\end{eqnarray}

The Jastrow factor can be formally represented in a many-body expansion
\begin{eqnarray}
  J = \sum_{\sigma i}u_1(\mathbf{r}_{\sigma i})+\frac{1}{2}\sum_{{\sigma\sigma'ij}}u_2(\mathbf{r}_{\sigma i},\mathbf{r}_{\sigma' j})+\frac{1}{6}\sum_{{\sigma\sigma'\sigma''ijk}}u_3(\mathbf{r}_{\sigma i},\mathbf{r}_{\sigma' j},\mathbf{r}_{\sigma'' k})+\cdots
\end{eqnarray}
with each $n$-body term $u_n$ being symmetric under particle
exchange. 

The one-body term is approximated in \qmcpack as a sum over atom-centered s-wave type functions that depend on the local ionic species $I$
\begin{eqnarray}
  u_1(\mathbf{r}_{\sigma i}) = \sum_{I\mu}u_{\sigma I}(\abs{\mathbf{r}_{\sigma i}-\mathbf{r}_{I\mu}})
\end{eqnarray}
with $r_{I\mu}$ being the position of the $\mu$-th ion of species $I$.  The dependence on spin is optional. 

The two-body term is approximated as a spin-dependent liquid-like factor (the electron-electron term) optionally with a second factor that additionally depends on the ionic coordinates (the electron-electron-ion term)
\begin{eqnarray}
  u_2(\mathbf{r}_{\sigma i},\mathbf{r}_{\sigma' j}) &= u_{\sigma\sigma'}(\abs{\mathbf{r}_{\sigma i}-\mathbf{r}_{\sigma' j}}) \\ 
  &+  \sum_{I\mu}u_{\sigma\sigma'I}(\abs{\mathbf{r}_{\sigma i}-\mathbf{r}_{I\mu}},\abs{\mathbf{r}_{\sigma' j}-\mathbf{r}_{I\mu}},\abs{\mathbf{r}_{\sigma i}-\mathbf{r}_{\sigma' j}}) \nonumber
\end{eqnarray}   
In each case, the up-up and down-down terms are constrained to be equal.

A wide range of options are available for the one-dimensional electron-ion ($u_{\sigma I}$) and electron-electron ($u_{\sigma\sigma'}$) Jastrow correlation functions including B-splines, first and second-order Pad\'{e} functions, long and short ranged Yukawa functions, and various short-ranged functions suitable for model helium.  The most commonly used choice for either correlation function is a one-dimensional cubic B-spline 
\begin{eqnarray}
  u(r) = \sum_{m=0}^Mp_mB_3\left(\frac{r}{r_c/M}-m\right)
\end{eqnarray}
where $B_3(x)$ denotes a cardinal cubic B-spline function defined on the interval $x\in [-3,1)$ (centered at $x=-1$), $\{p_n\}$ are the control points, and $r_c$ is the cutoff radius.  The last $M$ control points ($p_1\ldots p_M$) comprise the optimizable parameters while $p_0$ is determined by the cusp condition 
\begin{eqnarray}
p_0=p_2-\frac{2M}{r_c}\frac{\partial u}{\partial r}\bigg|_{r=0}
\end{eqnarray}
The Jastrow cutoffs should be selected in the region of non-vanishing
density in open boundary conditions. In periodic boundary conditions the cutoffs must be smaller than the simulation cell Wigner-Seitz radius.

The three-body electron-electron-ion correlation function ($u_{\sigma\sigma'I}$) currently used in \qmcpack is identical to the one proposed in \cite{Drummond2004}:  
\begin{eqnarray}
u_{\sigma\sigma'I}(r_{\sigma I},r_{\sigma'I},r_{\sigma\sigma'}) &= \sum_{\ell=0}^{M_{eI}}\sum_{m=0}^{M_{eI}}\sum_{n=0}^{M_{ee}}\gamma_{\ell mn} r_{\sigma I}^\ell r_{\sigma'I}^m r_{\sigma\sigma'}^n \\
   &\times \left(r_{\sigma I}-\frac{r_c}{2}\right)^3 \Theta\left(r_{\sigma I}-\frac{r_c}{2}\right) \nonumber \\
   &\times \left(r_{\sigma' I}-\frac{r_c}{2}\right)^3 \Theta\left(r_{\sigma' I}-\frac{r_c}{2}\right) \nonumber 
\end{eqnarray}
Here $M_{eI}$ and $M_{ee}$ are the maximum polynomial orders of the
electron-ion and electron-electron distances, respectively,
$\{\gamma_{\ell mn}\}$ are the optimizable parameters (modulo
constraints), $r_c$ is a cutoff radius, and $r_{ab}$ are the distances
between electrons or ions $a$ and $b$. i.e. The correlation function
is only a function of the interparticle distances and not a more
complex function of the particle positions, $\mathbf{r}$. As indicated by the
$\Theta$ functions, correlations are set to zero beyond a distance of
$r_c/2$ in either of the electron-ion distances and the largest
meaningful electron-electron distance is $r_c$.  This is the
highest-order Jastrow correlation function currently implemented.

Today, solid state applications of \qmcpack usually utilize one and
two-body B-spline Jastrow functions, with calculations on heavier
elements often also using the three-body term described above. While
there are not yet any comprehensive comparisons between the different
forms of the Jastrow factor in current use, this choice appears to
give very similar accuracy to other forms. Experience with atoms and
molecules is similar. In the future, should systematic studies find a new
form of Jastrow factor to be more efficient or effective, it
can be rapidly introduced due to the object oriented nature of the
application.


\section{Hamiltonian}
\label{sec:hamiltonian}

The Hamiltonian is represented in \qmcpack as a sum of abstract 
components
\begin{eqnarray}
  \hat{H} = \sum_n \hat{H}_n
\end{eqnarray}
with each component implemented as a class.  The functionality of all 
Hamiltonian component classes is dictated by a shared base class.  The 
primary shared characteristic of each component is the evaluation of 
its contribution to the local energy
\begin{eqnarray}
  E_L = \sum_nE_{Ln}\quad,\quad E_{Ln}\equiv \Psi_T^{-1}\hat{H}_n\Psi_T
\end{eqnarray}
In QMC algorithms, the local energy (as well as other observables) 
is collected after all Monte Carlo walkers have advanced one step in 
configuration space.

Possibly uniquely, the Hamiltonian that is solved is specified in the
\qmcpack input. This makes \qmcpack suitable for model
studies as well as ab initio calculations.
The most general Hamiltonian that can currently be handled by \qmcpack 
is non-relativistic with pairwise interactions between quantum 
(electrons or nuclei) or classical (nuclei only) particles and 
possibly external fields
\begin{eqnarray}
  \hat{H} = \sum_q\hat{T}_q + \sum_{q\ne q'}\hat{V}_{qq'} + \sum_{qc}\hat{V}_{qc} + \sum_{c\ne c'}V_{cc'} + \sum_q\hat{V}^{ext}_q + \sum_cV^{ext}_c
\end{eqnarray}
Here $q$ and $c$ denote the species of quantum and classical 
particles, respectively.

While non-adiabatic (multiple quantum species) and model-potential 
(e.g. low-temperature helium) calculations are possible, we focus
the remainder of the discussion to the most typical case: electronic 
structure problems within the Born-Oppenheimer (clamped nuclei) 
approximation \cite{born1927}.  In this case, the many body Hamiltonian is (in atomic 
units)
\begin{eqnarray}
  \hat{H} = -\frac{1}{2}\sum_i\nabla_i^2 + \sum_{i<j}v_{ee}(\mathbf{r}_i,\mathbf{r}_j) + \sum_i\sum_{I\mu}v_{eI}(\mathbf{r}_i,\mathbf{r}_{I\mu}) + \sum_{I\mu I'\mu'}v_{II'}(\mathbf{r}_{I\mu},\mathbf{r}_{I'\mu'})
\end{eqnarray}
where $i$ and $j$ sum over electron indices and $I\mu$ denotes the 
$\mu$-th ion with species $I$.

\qmcpack supports all-electron and pseudopotential calculations in both 
open and periodic boundary conditions.  The choice of ion core and 
boundary conditions affects the potential terms and we now briefly 
review these forms.  For all-electron calculations in open boundary 
conditions, all of the interaction potential terms are related simply 
to the bare Coulomb interaction $v_C(r) \equiv 1/r$
\begin{eqnarray}
  v_{ee}(\mathbf{r}_i,\mathbf{r}_j) &= v_C(\abs{\mathbf{r}_i-\mathbf{r}_j}) \\
  v_{eI}(\mathbf{r}_i,\mathbf{r}_{I\mu}) &= -Z_I v_C(\abs{\mathbf{r}_i-\mathbf{r}_{I\mu}}) \nonumber \\
  v_{II'}(\mathbf{r}_{I\mu},\mathbf{r}_{I'\mu'}) &= Z_IZ_{I'}v_C(\abs{\mathbf{r}_{I\mu}-\mathbf{r}_{I'\mu'}}) \nonumber
\end{eqnarray}

In periodic boundary conditions (PBC), the long-ranged part of each 
potential contributes an infinite number of terms due to the series 
of image cells filling all of space.  
\begin{eqnarray}
  v(\abs{\mathbf{r}_1-\mathbf{r}_2})\longrightarrow\sum_{n=0}^\infty v(\abs{\mathbf{r}_1-\mathbf{r}_2+n^{T} \mathbf{L}})
\end{eqnarray}
Sums of this type are evaluated via the Ewald summation technique
\cite{ewald1921}.  An optimized breakup \cite{natoli1995}
into long and short-ranged contributions is used to minimize
computational effort.

With the introduction of semi-local pseudopotentials, the electron-ion 
term takes the form
\begin{eqnarray}
  v_{eI}(\mathbf{r}_i,\mathbf{r}_{I\mu}) &= v_{loc}(\abs{\mathbf{r}_i-\mathbf{r}_{I\mu}}) + \sum_{\ell m}\ket{Y_{\ell m}} \left[ v_\ell(\abs{\mathbf{r}_i-\mathbf{r}_{I\mu}}) - v_{loc}(\abs{\mathbf{r}_i-\mathbf{r}_{I\mu}}) \right]\bra{Y_{\ell m}}
\end{eqnarray}
where all of the non-local channel terms vanish beyond cutoff radii
that may be unique to each channel and the local part approaches
$-Z_{eff}/r$ in the long distance limit ($Z_{eff}$ is the effective
core charge presented by the pseudopotential).  The evaluation of the
local energy for semi-local pseudopotentials follows the algorithm
laid out by Mitas {\it et al.}\cite{Mitas1991} with a 12 point angular
integration used by default.

In DMC calculations, the semi-local potentials are
evaluated within the locality approximation \cite{Mitas1991},
or the more recent ``t-moves'' approximations\cite{Casula2006,Casula2010} that restore the
variational principle the the DMC algorithm. In particular,
the algorithm of \cite{Casula2010} restores size-extensivity.


\section{Boundary conditions}
\label{sec:boundaryconditions}
\qmcpack accommodates both periodic and open boundary conditions in 1, 2, or 3 dimensions, including mixed boundary conditions. After the pseudopotential and fixed-node approximations in QMC, the choice of boundary conditions imposes another set of approximations onto a system that must be treated with care.

\subsection{Long-Range Interactions}
The long-ranged Coulombic interactions of the electrons and ions must be handled with care in order to ensure that the potential energy doesn't diverge when using periodic boundary conditions.  In \qmcpack, the interparticle interactions are computed using an optimized implementation\cite{natoli1995} of the well-known strategy of decomposing the interactions into short and long ranged components, and performing sums over the former and latter in real and reciprocal space, respectively \cite{ewald1921}.

\subsection{Twist-averaged boundary conditions}
Bloch's theorem demonstrates how a finite wavefunction can be used to simulate an infinite lattice within periodic boundary conditions by incorporating the following symmetry:
\begin{equation}
	\Psi(\mathbf{r}_{1} + \mathbf{L_{m}}, \mathbf{r}_{2}, \dots, \mathbf{r}_{N}) = e^{i\mathbf{K}\cdot\mathbf{L}_{m}} \Psi(\mathbf{r}_{1}, \mathbf{r}_{2}, \dots, \mathbf{r}_{N}) 
\end{equation}
where $\mathbf{K}$ is a vector in reciprocal space, $\mathbf{L}_{m}$ is a lattice vector of the supercell, and $\Theta_{m} \equiv \mathbf{K} \cdot \mathbf{L}_{m}$, is the ``twist angle'' \cite{martin_interacting_2016}. For pure periodic boundary conditions (in which $\Theta=0$), systems converge slowly to their thermodynamic limit due to shell effects and quantization of momentum \cite{Lin2001}. Therefore, to improve convergence speed and accuracy, one should average over many simulations done with different twist angles, a scheme called ``twist-averaged boundary conditions''. In \qmcpack, the averaging is done in post-processing, using e.g. \texttt{qmca} (section \ref{sec:averaging}) and/or Nexus (section \ref{sec:nexus}).


\section{Optimization}
\label{sec:optimization}

In all real-space quantum Monte Carlo calculations, optimizing the
wavefunction significantly improves both the accuracy and efficiency
of computation.  However, it is very difficult to directly adopt
deterministic minimization approaches due to the stochastic nature of
quantities with Monte Carlo.  Thanks to major algorithmic
improvements, it is now feasible to optimize up to tens of thousands
of parameters in a wavefunction for a large solid or molecule.
\qmcpack implements multiple optimizers based on the state-of-the-art
linear method with several techniques, described below, improving its
robustness, efficiency and capability.

\subsection{The Linear Method}

\qmcpack optimizes trial functions using an implementation of the linear method (LM)
\cite{UmrTouFilSorHen-PRL-07}
that includes modifications to
improve stability in the face of variables of greatly differing stiffnesses,
facilitate the optimization of excited states,
and reducing the memory footprint when optimizing large numbers of
variational parameters. The LM is sometime referred to as ``energy
minimization'', although the approach is more general.
The LM gets its name from the way that it employs a linear expansion of the wavefunction,
\begin{eqnarray}
\label{eqn:lm_wfn}
|\Phi\rangle = \sum_{y=0}^{N_\mathrm{v}} c_y |\Psi^y\rangle,
\end{eqnarray}
where $|\Psi^y\rangle$ for $y\in\{1,2,3,...\}$ is the derivative of the trial function $|\Psi\rangle$ with respect to its $y$th variational parameter and $|\Psi^0\rangle\equiv|\Psi\rangle$,
within an expanded energy expression,
\begin{eqnarray}
\label{eqn:lm_expanded_energy}
E(\hspace{0.2mm}\vec{c}\hspace{0.6mm})
=\frac{\langle\Phi|H|\Phi\rangle}{\langle\Phi|\Phi\rangle}.
\end{eqnarray}
Using this linear approximation to how the energy changes with the variational parameters,
minimizing $E$ with respect to $\vec{c}$ can be achieved by solving the generalized eigenvalue problem
\begin{eqnarray}
\label{eqn:lm_gev}
\sum_{y=0}^{N_\mathrm{v}} \frac{\langle\Psi^x|H|\Psi^y\rangle}{\langle\Psi|\Psi\rangle} c_y
= E \sum_{y=0}^{N_\mathrm{v}} \frac{\langle\Psi^x|\Psi^y\rangle}{\langle\Psi|\Psi\rangle} c_y
\end{eqnarray}
or, written in matrix-vector notation,
\begin{eqnarray}
\label{eqn:lm_gev_mv}
\mathbf{H} \hspace{1mm} \vec{c} = E \hspace{1mm} \mathbf{S} \hspace{1mm} \vec{c},
\end{eqnarray}
the matrix elements for which are evaluated by Monte Carlo integration \cite{TouUmr-JCP-07,TouUmr-JCP-08}
in direct analogy to how VMC evaluates the energy. 
If one assumes the improved trial function $|\Phi\rangle$ is similar to the previous
trial function $|\Psi\rangle$, which implies that the ratio $c_x/c_0$ is small for all $x\in\{1,2,3,...\}$,
then a reasonable approximation to $|\Phi\rangle$ can be had by replacing
$\mu_x\rightarrow\mu_x+c_x/c_0$ for each variational parameter
$\mu_x$ in $|\Psi\rangle$.
As for other optimization methods that compute an update based on some local
approximation to the target function, such as Newton-Raphson, this process is then repeated until further updates no longer
lower the energy.

\subsection{Stabilizing the Linear Method}

In practice, it is important to implement an analogue to the trust radius schemes common to Newton-Raphson
in order to ensure that the solution of equation\ (\ref{eqn:lm_gev}) does not correspond to an unreasonably
long step in variable space, or, put another way, to ensure that the ratio $c_x/c_0$ is not too large.
The LM optimizer in \qmcpack supports two mechanisms for preventing too-large updates: a diagonal
shift $\alpha$ as employed in the original algorithm \cite{UmrTouFilSorHen-PRL-07} as well as an overlap-based
shift $\beta$ that becomes important when parameters of greatly different stiffnesses are present.
Using these shifts, the Hamiltonian matrix is modified to become
\begin{eqnarray}
\label{eqn:lm_shifted_ham}
\mathbf{H} \rightarrow \mathbf{H} + \alpha \mathbf{A} + \beta \mathbf{B},
\end{eqnarray}
where $\mathbf{A}$ and $\mathbf{B}$ provide stabilization via the original and overlap
shifts, respectively.
As in the original method, \qmcpack uses $A_{xy}=\delta_{xy}(1-\delta_{x0})$ and the adjustable shift strength
$\alpha$ to effectively raise the energy along each direction of change while
leaving the current wavefunction $|\Psi^0\rangle=|\Psi\rangle$ unaffected.

While the original shifting scheme has been effective in many cases, it can struggle if two different
variational parameters produce wavefunction derivatives of vastly different sizes.
For example, imagine a two-variable wavefunction whose overlap matrix evaluates to
\begin{eqnarray}
\label{eqn:lm_bad_overlap}
\mathbf{S} =
\left[
\begin{array}{c c c} 1 \hspace{1.5mm} & 0 & 0 \\ 0 & 1 & 0 \\ 0 & 0 & 10^6 \end{array}
\right].
\end{eqnarray}
Performing the usual
$\vec{\nu}=(\mathbf{S}^{\frac{1}{2}})\vec{c}$
 transformation to produce a standard eigenvalue problem
(with $\beta$ set to zero for now)
gives us
\begin{eqnarray}
\label{eqn:lm_innefective_shift}
\mathbf{S}^{-\frac{1}{2}}
\hspace{0.6mm}
\mathbf{H}
\hspace{0.6mm}
\mathbf{S}^{-\frac{1}{2}}
\hspace{0.6mm}
\vec{\nu}
\hspace{0.4mm}
+ 
\left[
\begin{array}{c c c}
1 & 0 & 0 \\
0 & \alpha & 0 \\
0 & 0 & \frac{\alpha}{10^6} \\
\end{array}
\right]
\vec{\nu} = E \hspace{0.3mm}\vec{\nu}.
\end{eqnarray}
We see that, if we were to make $\alpha$ large enough to significantly penalize the second variable direction,
we would penalize the first direction so much that it would essentially become a fixed parameter.

The purpose of the overlap shift is to resolve this issue by adding an energy penalty based on the norm of the
part of $\vec{c}$ corresponding to directions orthogonal to the current wavefunction $|\Psi\rangle$, which would
correctly penalize steps along directions of large derivative norms more than those along directions of small
derivative norms.
This goal is accomplished by the definitions
\begin{eqnarray}
\label{eqn:lm_overlap_shift_matrices}
Q_{xy} &= \delta_{xy} - \delta_{x0}(1-\delta_{y0}) S_{0y} \\
T_{xy} &= (1 - \delta_{x0}\delta_{y0}) \left[\mathbf{Q}^+\mathbf{S}\hspace{0.4mm}\mathbf{Q}\right]_{xy} \\
\mathbf{B} &= \left(\mathbf{Q}^+\right)^{-1} \hspace{0.1mm} \mathbf{T} \hspace{0.6mm} \mathbf{Q}^{-1}
\end{eqnarray}
in which $\mathbf{Q}$ transforms into a basis in which all update directions are orthogonal to
the current wavefunction $|\Psi\rangle$ (this transformation is equivalent to that
of equation\ (24) of reference \cite{TouUmr-JCP-07}).
$\mathbf{T}$ is the overlap matrix in this basis with its first element zeroed out so that the
current wavefunction is not penalized.
Finally, the inverses of $\mathbf{Q}$ and its adjoint transform us back to the basis of the
original generalized eigenvalue problem so that the effect of the overlap shift cay
be written in the form of equation\ (\ref{eqn:lm_shifted_ham}).
Note that, in practice, it is not necessary to construct $\mathbf{B}$ explicitly, as \qmcpack
solves the generalized eigenvalue equation by iterative Krylov subspace expansion, during
which the Krylov basis (whose first vector is always $|\Psi\rangle$) is kept orthonormal
by the Gram-Schmidt procedure.
In this Krylov basis, applying the overlap shift involves merely adding $\beta$ to the
diagonal of the subspace Hamiltonian matrix (except, of course, to the first element
corresponding to $|\Psi\rangle$).
This Krylov approach also has the benefit of ensuring that the overall update is
orthogonal to the current wavefunction, which is related to norm-conservation and
was found to be desirable by the LM's original developers.
\cite{UmrTouFilSorHen-PRL-07,TouUmr-JCP-07,TouUmr-JCP-08}

Although like most trust-radius schemes the optimal choices for $\alpha$ and $\beta$ are
somewhat heuristic, \qmcpack automatically adjusts them after each iteration
of the LM by solving for the updates generated by three different sets of shifts and
retaining the shift that gave the best update, as determined by a correlated-sampling
comparison of their energies on a fresh sample.
For maximum efficiency in regimes where optimization is not difficult but sampling is
expensive, \qmcpack retains the ability to run in a single-shift, no-second-sample mode. 
When running instead in multi-shift mode, we have observed that successful optimizations
often result with the simple initial choice of $\alpha=\beta=1$.
In principle, however, one might expect $\alpha<\beta$ to be more effective, because when
the $\beta$ shift is filling the role of limiting the update size, $\alpha$ is only
needed to penalize (hopefully rare) linear dependencies between update directions
that $\beta$, being overlap-based, cannot address.

\subsection{Optimizing for Excited States}

\qmcpack's current LM optimization engine supports both standard energy minimization
and the minimization of a recently introduced \cite{Zhao:2016:dir_tar} excited state
target function,
$\langle\Psi|(\omega-\hat{H})|\Psi\rangle / \langle\Psi|(\omega-\hat{H})^2|\Psi\rangle$,
whose global minimum is the exact energy eigenstate immediately above
the targeted energy $\omega$.
Although this technology is a very recent development and will doubtless evolve
in time as the science behind excited state targeting matures, we felt it important
to make an early version of it available to the community.
Optimization proceeds in much the same way as for a ground state, with the user
specifying $\omega$ and the stabilization shifts $\alpha$ and $\beta$ and the
LM repeatedly solving generalized eigenvalue equations analogous to
equation\ (\ref{eqn:lm_gev}) to generate wavefunction updates.
Additional methods for automatically selecting and updating $\omega$ have been developed\cite{Shea:2017:size_con_excited_states}.
For details into this targeting function and how it is optimized, we refer the
reader to the original publication \cite{Zhao:2016:dir_tar}.

\subsection{Handling Large Parameter Sets}

One important limitation of the LM comes when the number of variational parameters
rises to 10,000 or more, at which point the contributions to $\mathbf{H}$ and $\mathbf{S}$
made by each Markov chain become cumbersome to store in memory, especially
when running one Markov chain per core on a large parallel system in which per-core memory
is limited.
\qmcpack currently addresses this memory bottleneck using the blocked LM\cite{zhao_blocked_2017},
a recent algorithm that separates the variable space into blocks,
estimates the most important variable-change directions within each block,
and then uses these directions to construct a reduced and
vastly more memory efficient LM eigenvalue problem to generate an update
direction in the overall variable space.
Like excited state targeting, this is a new feature that can be expected to
evolve in time, and has been made openly available to the community in the
spirit of rapid dissemination.
As of this writing, it has not been widely tested outside of the work in
its original publication \cite{zhao_blocked_2017}, but in time we expect
to have a clearer picture of its capabilities.

\subsection{Multi-objective optimization}
\qmcpack also supports optimizing variational parameters based on not
only the total energy but also variance. 
In certain situations, the best target object may not be the energy only but a cost function mixing both energy and variance
which reduces to zero when the wavefunction is exact. The cost
function can be any linear combination of energy and variance.
\qmcpack picks the optimal parameter set corresponding to the minimal value of a quartic function
fitting the cost function evaluated on seven shifts by correlated-sampling.


\section{Observables}
A broad range of observables and estimators are available in \qmcpack.
In this section we describe the total number density (density), number
density resolved by particle spin (spindensity), spherically averaged
pair correlation function (gofr), static structure factor (sk), energy
density (energydensity), one body reduced density matrix (dm1b) and
force (Forces) estimators. These estimators can be 
evaluated for the entire run (e.g. all VMC and DMC sections) when added to the
Hamiltonian section in the input file, or applied to a specific
section. Higher order density matrix quantities for calculating quantum entanglement have also been studied previously, e.g. \cite{mcminis-entanglement-2013,swingle-entanglement-2013,tubman-entanglement-2014}.

\subsection{Density and spin density}
\begin{figure}
\begin{center}
\includegraphics[trim = 0mm 0mm 0mm 0mm, clip,width=0.8\columnwidth]{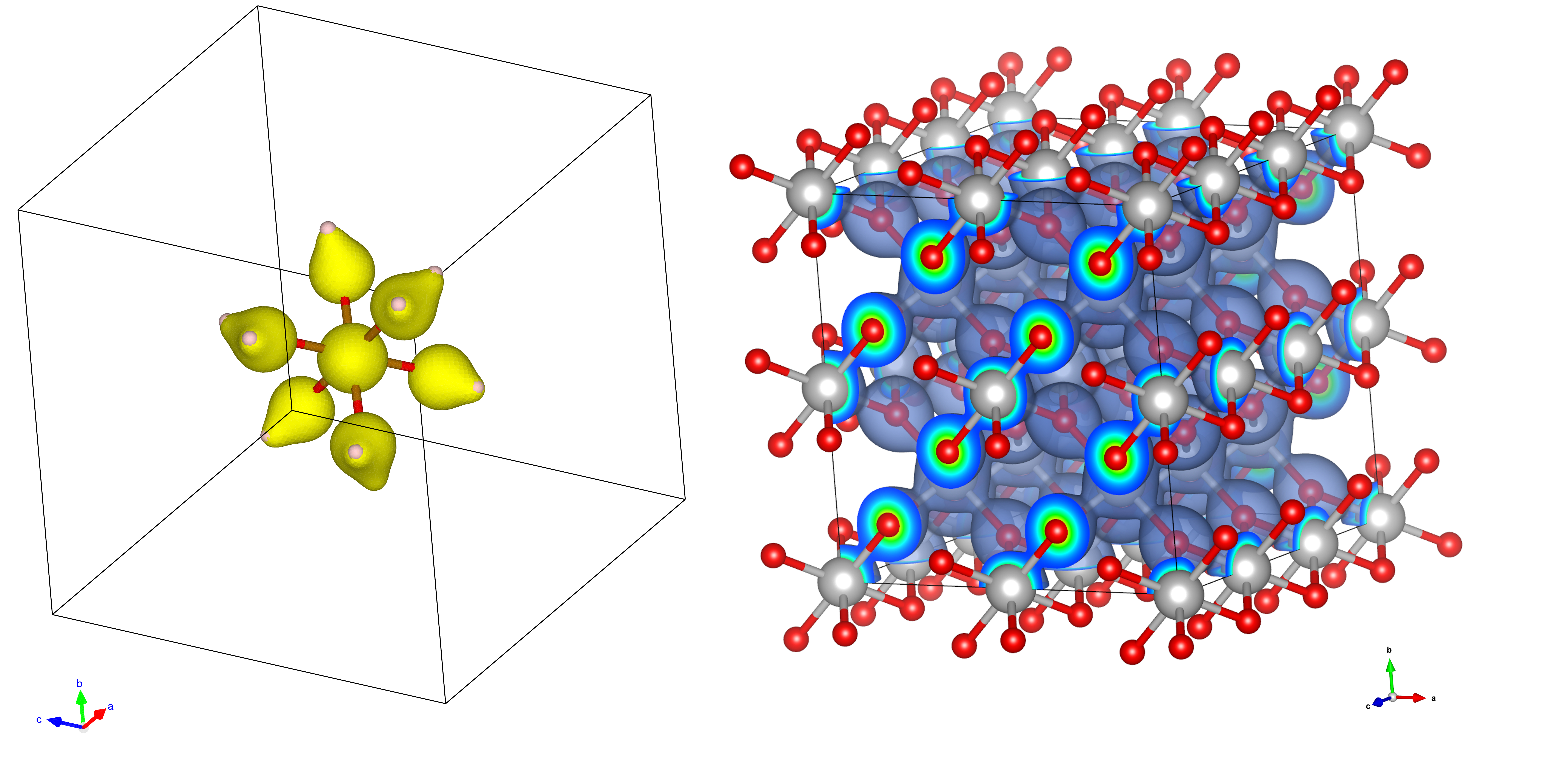}
\end{center}
\caption{Left: Electron density of $(Fe(H_2O)_6)^{2+}$ cluster in a
  10\AA~ box using 100x100x100 grid points, based on calculations in \cite{SongUnpub2018}. Right:
  Electron density of $TiO_2$ in a grid of 80x80x80 grid
  points, based on calculations in \cite{Luo2016NJP}.
\label{fig:Densities}
}
\end{figure}

The particle number density operator is given by
\begin{equation}
\hat{n}_r=\sum_{i} \delta (\mathbf{r}-\mathbf{r}_i), 
\end{equation}
This estimator accumulates the number density on a uniform histogram grid over the simulation cell. The value obtained for a grid cell \textit{c} with volume $\Omega_c$ is then the average number of particles in that cell
\begin{equation}
n_c=\int{} d\mathbf{R}|\Psi|^2\int{_{\Omega_c}}d\mathbf{r}\sum_i\delta(\mathbf{r}-\mathbf{r}_i), 
\end{equation}
When using periodic boundary conditions, the density will be collected for the cell (or supercell) defined by the simulation. When using non-periodic boundary conditions, a cell has to be defined in order to set a grid. It is then recommended to center the system (molecule) in the middle of the defined cell. The collected data is stored in HDF5 format in a \textit{.stat.h5} file. Using Nexus (section \ref{sec:nexus}), one can use the qdens tool to extract the data in a \textit{*.xsf} format readable with visualization tools such as XCrysDens\cite{Kokalj1999176,Kokalj2003155} or VESTA\cite{Momma:db5098}. Examples of density plots are shown in figure~\ref{fig:Densities}. Similar to the density, the spin-density estimator can be also collected for each independent spin, as shown and analyzed in \cite{Benali2016PCCP} for magnetic states in $Ti_4O_7$.
 
\subsection{Pair correlation function}

The functional form of the species-resolved radial pair correlation function operator is
\begin{eqnarray}
  g_{ss'}(r) = \frac{V}{4\pi r^2N_sN_{s'}}\sum_{i_s=1}^{N_s}\sum_{j_{s'}=1}^{N_{s'}}\delta(r-|\mathbf{r}_{i_s}-\mathbf{r}_{j_{s'}}|).
\end{eqnarray}
Here $N_s$ is the number of particles of species $s$ and $V$ is the supercell volume.  If $s=s'$, then the sum is restricted so that $i_s\ne j_s$.

An estimate of $g_{ss'}(r)$ is obtained as a radial histogram with a set of $N_b$ uniform bins of width $\delta r$.  This can be expressed analytically as
\begin{eqnarray}
  \tilde{g}_{ss'}(r) = \frac{V}{4\pi r^2N_sN_{s'}}\sum_{i=1}^{N_s}\sum_{j=1}^{N_{s'}}\frac{1}{\delta r}\int_{r-\delta r/2}^{r+\delta r/2}dr'\delta(r'-|\mathbf{r}_{si}-\mathbf{r}_{s'j}|),
\end{eqnarray}
where the radial position $r$ is restricted to reside at the bin centers  $\delta r/2, 3 \delta r/2, \ldots$.

\subsection{Static structure factor}
Let $\rho^e_{\mathbf{k}}=\sum_j e^{i \mathbf{k}\cdot\mathbf{r}_j^e}$ be the Fourier space electron density, with $\mathbf{r}^e_j$ being the coordinate of the j-th electron.  $\mathbf{k}$ is a wavevector commensurate with the simulation cell.  The static electron structure factor $S(\mathbf{k})$ can be measured at all commensurate $\mathbf{k}$ such that $|\mathbf{k}| \leq (\mathrm{LR\_DIM\_CUTOFF}) r_c$.  $N^e$ is the number of electrons, \texttt{LR\_DIM\_CUTOFF} is the optimized breakup parameter, and $r_c$ is the Wigner-Seitz radius.  It is defined as follows:
\begin{equation}
S(\mathbf{k}) = \frac{1}{N^e}\langle \rho^e_{-\mathbf{k}} \rho^e_{\mathbf{k}} \rangle
\end{equation}

\subsection{Energy density estimator}
An energy density operator, $\hat{\mathcal{E}}_r$,  satisfies
\begin{eqnarray}
  \int dr \hat{\mathcal{E}}_r = \hat{H},
\end{eqnarray}
where the integral is over all space and $\hat{H}$ is the Hamiltonian.  In \qmcpack, the energy density is split into kinetic and potential components
\begin{eqnarray}
  \hat{\mathcal{E}}_r = \hat{\mathcal{T}}_r + \hat{\mathcal{V}}_r 
\end{eqnarray}
with each component given by
\begin{eqnarray}
   \hat{\mathcal{T}}_r &=  \frac{1}{2}\sum_i\delta(\mathbf{r}-\mathbf{r}_i)\hat{p}_i^2 \\  
   \hat{\mathcal{V}}_r &= \sum_{i<j}\frac{\delta(\mathbf{r}-\mathbf{r}_i)+\delta(\mathbf{r}-\mathbf{r}_j)}{2}\hat{v}^{ee}(\mathbf{r}_i,\mathbf{r}_j)
              + \sum_{i\ell}\frac{\delta(\mathbf{r}-\mathbf{r}_i)+\delta(\mathbf{r}-\tilde{\mathbf{r}}_\ell)}{2}\hat{v}^{eI}(\mathbf{r}_i,\tilde{\mathbf{r}}_\ell) \nonumber\\ 
    &\qquad   + \sum_{\ell< m}\frac{\delta(\mathbf{r}-\tilde{\mathbf{r}}_\ell)+\delta(\mathbf{r}-\tilde{\mathbf{r}}_m)}{2}\hat{v}^{II}(\tilde{\mathbf{r}}_\ell,\tilde{\mathbf{r}}_m).\nonumber
\end{eqnarray}
Here $\mathbf{r}_i$ and $\tilde{\mathbf{r}}_\ell$ represent electron and ion positions, respectively, $\hat{p}_i$ is a single electron momentum operator, and $\hat{v}^{ee}(\mathbf{r}_i,\mathbf{r}_j)$, $\hat{v}^{eI}(\mathbf{r}_i,\tilde{\mathbf{r}}_\ell)$, $\hat{v}^{II}(\tilde{\mathbf{r}}_\ell,\tilde{\mathbf{r}}_m)$ are the electron-electron, electron-ion, and ion-ion pair potential operators (including non-local pseudopotentials, if present).  This form of the energy density is size-consistent, \textit{i.e.} the partially integrated energy density operators of well separated atoms gives the isolated Hamiltonians of the respective atoms.  For periodic systems with twist averaged boundary conditions, the energy density is formally correct only for either a set of supercell k-points that correspond to real-valued wavefunctions, or a k-point set that has inversion symmetry around a k-point having a real-valued wavefunction.  For more information about the energy density, see \cite{Krogel2013}.

The energy density can be accumulated on piecewise uniform three dimensional grids in generalized Cartesian, cylindrical, or spherical coordinates.  The energy density integrated within Voronoi volumes centered on ion positions is also available.  The total particle number density is also accumulated on the same grids by the energy density estimator for convenience so that related quantities, such as the regional energy per particle, can be computed easily.

\subsection{One-body density matrix}
The N-body density matrix in DMC is $\hat{\rho}_N=\operator{\Psi_{T}}{}{\Psi_{FN}}$
 (for VMC, substitute $\Psi_T$ for $\Psi_{FN}$).  The one body reduced density matrix (1RDM) is obtained by tracing out all particle coordinates but one:
\begin{eqnarray}
  \hat{n}_1 &= \sum_nTr_{R_n}\operator{\Psi_{T}}{}{\Psi_{FN}}
\end{eqnarray}
In the formula above, the sum is over all electron indices and $Tr_{\mathbf{R}_n}(*)\equiv\int d\mathbf{R}_n\expval{\mathbf{R}_n}{*}{\mathbf{R}_n}$
 with $\mathbf{R}_n=[\mathbf{r}_1,...,\mathbf{r}_{n-1},\mathbf{r}_{n+1},...,\mathbf{r}_N]$.  When the sum is restricted over spin up or down electrons, one obtains a density matrix for each spin species.  The 1RDM computed by \qmcpack is partitioned in this way.

In real space, the matrix elements of the 1RDM are
\begin{eqnarray}
  n_1(\mathbf{r},\mathbf{r}') &= \expval{\mathbf{r}}{\hat{n}_1}{\mathbf{r}'} = \sum_n\int d\mathbf{R}_n \Psi_T(\mathbf{r},\mathbf{R}_n)\Psi_{FN}^*(\mathbf{r}',\mathbf{R}_n) 
\end{eqnarray}

A more efficient and compact representation of the 1RDM is obtained by
expanding in single particle orbitals, e.g. from a Hartree-Fock or DFT calculation, $\{\phi_i\}$:
\begin{eqnarray}\label{eq:dm1b_direct}
  n_1(i,j) &= \expval{\phi_i}{\hat{n}_1}{\phi_j} \nonumber \\
           &= \int d\mathbf{R} \Psi_{FN}^*(\mathbf{R})\Psi_{T}(\mathbf{R}) \sum_n\int d\mathbf{r}'_n \frac{\Psi_T(\mathbf{r}_n',\mathbf{R}_n)}{\Psi_T(\mathbf{r}_n,\mathbf{R}_n)}\phi_i(\mathbf{r}_n')^* \phi_j(\mathbf{r}_n) 
\end{eqnarray} 

The integration over $\mathrm{r}'$ in equation \ref{eq:dm1b_direct} is inefficient when one is also interested in obtaining matrices involving energetic quantities, such as the energy density matrix \cite{Krogel2013} or the related and more well known Generalized Fock matrix.  For this reason, we compute:\cite{Krogel2013} 
\begin{eqnarray}
    n_1(i,j) \approx \int d\mathbf{R} \Psi_{FN}(\mathbf{R})^*\Psi_T(\mathbf{R})  \sum_n \int d\mathbf{r}_n' \frac{\Psi_T(\mathbf{r}_n',\mathbf{R}_n)^*}{\Psi_T(\mathbf{r}_n,\mathbf{R}_n)^*}\phi_i(\mathbf{r}_n)^* \phi_j(\mathbf{r}_n') 
\end{eqnarray}

\subsection{Forces}
For all-electron calculations, na\"ively estimating the bare Coulomb Hellman-Feynman force with Quantum Monte Carlo suffers from a fatal problem: While the expectation value of this estimator is well defined, the $1/r^2$ divergence causes the variance to be infinite, meaning we can't obtain a meaningful error bar for this quantity.  There are several schemes to circumvent this.  For all-electron calculations, \qmcpack can currently calculate forces and stress using the Chiesa estimator \cite{Chiesa2005} in both open and periodic boundary conditions.  Implementation details and validation of forces in periodic boundary conditions can be found in \cite{Clay2016}.  In the future, pseudopotential forces will be supported, and methods to reduce the variance of existing estimators are currently being explored.

\section{Forward-Walking Estimators}
\label{sec:forward_walking}
Forward-walking is a method by which one can sample the pure fixed-node distribution $\langle \Phi_0 | \Phi_0\rangle$.  Specifically, one multiplies each walker's DMC mixed estimate for the observable $\mathcal{O}$, $\frac{\mathcal{O}(\mathbf{R})\Psi_T(\mathbf{R})}{\Psi_T(\mathbf{R})}$, by the weighting factor $\frac{\Phi_0(\mathbf{R})}{\Psi_T(\mathbf{R})}$.  This weighting factor for any walker $\mathbf{R}$ is proportional to the total number of descendants the walker will have after a sufficiently long projection time $\beta$.  

To forward-walk on an observable, one declares a generic forward-walking estimator within a Hamiltonian block, and then specifies the observables to forward-walk on and forward-walking parameters.



\section{Orbital space QMC methods}
\label{sec:orbital_methods}

\subsection{Introduction}
In addition to real-space QMC methods, \qmcpack also supports orbital-space QMC approaches
for the study of atomic, molecular and solid-state systems.  Auxiliary-Field Quantum Monte Carlo (AFQMC) is implemented internally, while interfaces to selected Configuration Interaction (SCI) methods have been developed.\cite{Nesbet1955,Caffarel2013,tubman-selected-ci-2016,holmes-selected-ci-2016} 
The starting point of orbital-space approaches 
is the Hamiltonian in second quantization, typically defined by
\begin{equation}
\hat{H}=\sum_{i,j} h_{ij} \hat{c}^{\dagger}_i \hat{c}_j + \frac{1}{2} \sum_{i,j,k,l} V_{ijkl}  \hat{c}^{\dagger}_i \hat{c}^{\dagger}_j  \hat{c}_l \hat{c}_k, 
\end{equation}
where $\hat{c}^{\dagger}_i (\hat{c}_i)$ are the creation (annihilation) operators for spinors associated with a given single particle basis set, with associated 1- and 2-electron matrix elements given by $h_{ij}$ and $V_{ijkl}$. The choice of the single particle basis along with the calculation of the appropriate Hamiltonian matrix elements must be performed by a separate electronic structure package. The Hamiltonian matrix elements are expected in the \textit{FCIDUMP} format used by codes including Molpro \cite{molpro}, PySCF \cite{pyscf2018} and VASP \cite{kresse_efficiency_1996,kresse_efficient_1996,kresse_ultrasoft_1999}. The calculations are typically performed on a single particle basis defined by the solution of a Hartree-Fock or DFT calculation. Both finite and periodic calculations are possible.

\subsection{Auxiliary Field Quantum Monte Carlo (AFQMC)}
The fundamental idea behind AFQMC is identical to that of DMC, namely that the propagation of many-body states in imaginary time leads to the lowest eigenstate of the Hamiltonian with non-zero overlap\cite{ZhangPRB1997}. In contrast with DMC, AFQMC operates in the Hilbert space of non-orthogonal Slater determinants and uses the Hubbard-Stratonovich transformation\cite{HUbbardPRL1959,HirschPRB1983} to express the short-time approximation of the propagator as an integral over propagators that contain only 1-body terms. The application of one-body propagators to walkers in the algorithm (represented by Slater determinants) leads to rotations of the corresponding Slater determinants that define a random walk, similar to the random walk in real-space followed by walkers in DMC\cite{HirschPRB1985}. \qmcpack implements the constrained-path algorithm of Zhang and Krakauer with the phaseless approximation \cite{ZhangPRB1997,Zhang03}. Similar to the algorithm of Zhang and Krakauer, we use importance sampling and force-bias to improve the sampling efficiency of the algorithm. For a complete description of the implemented algorithm, see the lecture notes on AFQMC by S. Zhang in the open-access book \cite{pavarini_emergent_2013} and \cite{Motta2017}.

The AFQMC implementation in \qmcpack attempts to minimize the memory requirements of the calculation, while increasing the performance of the associated computations. This is done by a combination of: (1) distributed sparse representations of large data structures (e.g. 2-electron integrals), (2) efficient use of shared-memory on multi-core architectures, (3) combination of efficient BLAS and sparse-BLAS routines for all major computations, and (4) an efficient distributed algorithm for walker propagation. Notice that the code is able to distribute the work associated with the propagation of a walker over many nodes, enabling access to systems with thousands of basis functions with a full ab initio representation. Both single determinant as well as multi-determinant trial wave-functions are implemented. In the case of multi-determinant expansions, both orthogonal as well as non-orthogonal expansions are efficiently implemented. For orthogonal expansions, a fast algorithm based on the Sherman-Morrison-Woodbury formula is implemented which leads to a modest increase in computing time for determinant expansions involving even many thousands of terms.

\subsection{Selected CI and CIPSI wavefunction interface}
\label{sec:cipsi}

As discussed previously, a direct path towards improving the accuracy of a QMC calculation is through a better trial wavefunction. One approach is to use Selected CI methods such as CIPSI (Configuration Interaction using a Perturbative Selection done Iteratively), or the recently developed Adaptive Sampling CI (ASCI)\cite{tubman-selected-ci-2016} and Heat Bath CI (HBCI)\cite{holmes-selected-ci-2016}. The principle behind selected CI methods was first published in 1955 by Nesbet\cite{Nesbet1955}. The first calculations on atoms were performed by Diner, Malrieu and Claverie\cite{Diner1967} in 1967.  Many advances have since been made with selected CI techniques, and it has been applied widely to atomic, molecular and periodic systems\cite{stampfub2005,harrison-cipsi-1991,sci-history-1,sci-history-2,sci-history-3,sci-history-4,zarea-selected-ci-2017,evan2016}. The method is based on an iterative process during which a wavefunction is improved at each step.  During each iteration, the current wavefunction is used in conjunction with the Hamiltonian to find important contributions that will be added to the wavefunction in the next iteration.  In most Selected CI approaches, the importance of a contribution is determined from a many body perturbation theory estimate. A full description of CIPSI, its algorithms, and results on various systems can be found in Refs. \cite{Caffarel2013,Scemama2018,Garniron2017-2}.  A description of new improvements to selected CI techniques that have been demonstrated with ASCI and HBCI can be found in Refs.~\cite{tubman-selected-ci-2016,holmes-selected-ci-2016}.  The CIPSI method\cite{Caffarel2013, Scemama2016,Scemama2018,Garniron2017-2} is implemented in the \textit{Quantum Package} (QP) code\cite{QP} developed by the Caffarel group. \qmcpack does not implement CIPSI, but is able to use output from the QP code via tight integration.

\begin{figure}
\begin{center}
\includegraphics[trim = 0mm 0mm 0mm 0mm, clip,width=0.25\columnwidth]{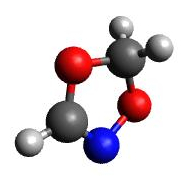}
\end{center}
\caption{The C$_2$O$_2$H$_3$N molecule.  The colors red, gray, blue and white correspond to oxygen, carbon, nitrogen, and hydrogen respectively.  The geometry is from the heterocyclic rings database\cite{Yu2015}.
\label{fig:C2O2H3N}
}
\end{figure}

In the following we use the C$_2$O$_2$H$_3$N molecule (figure \ref{fig:C2O2H3N}) to illustrate the use of CIPSI to obtain an improved trial wavefunction. The C$_2$O$_2$H$_3$N molecule is part of the cycloreversion of heterocyclic rings database\cite{Yu2015}, for which the geometry was optimized with DFT using the B3LYP function in a 6-31G basis set. Orbitals are represented within the aug-ccpVTZ basis set. The energetics of this molecule are known to have a strong dependence on the choice of functional in DFT simulations~\cite{Yu2015}. Diagnostics based on coupled cluster theory (CC) with single, double, and peturbative triple excitations (CCSD(T))\cite{szabo:book} suggest  a multireference character~\cite{Lee1989}, a known problem for these techniques \cite{lehtola-selected-ci-2017}. The multireference capability of DMC-CIPSI makes it an ideal tool for treating difficult systems with large static correlations.  

The FCI space for C$_2$O$_2$H$_3$N in aug-ccpVTZ is approximately $10^{88}$ determinants. Fortunately, using all of these determinants is not necessary to converge a QMC calculation to chemical accuracy. We truncate determinants based on their magnitudes with a user defined threshold $\epsilon$~\cite{Scemama2018}, which allows the wavefunction to be evaluated in QMCPACK with a cost growing as  $\sqrt[]{N_{det}}$, where $N_{det}$ is the number of determinants. Truncation values of $10^{-3}$, $10^{-4}$, $10^{-5}$ and $10^{-6}$ result in wavefunctions of 239, 44539, 541380 and 908128 determinants, respectively. For each truncated wavefunction we optimize one, two and three-body Jastrow factors with VMC. To isolate the improvement of the nodal surface when adding determinants from CIPSI, the coefficients of the determinants were not optimized, although this could result in a further improvement in the wavefunctions. DMC results are extrapolated to a zero time-step using time-steps of $\tau=0.001,~0.0008$ and $0.0005$.

\begin{table}[t]
\centering
\begin{tabular}{l|c|c|c}
\hline \hline
 N$_{det}$ &CIPSI(E) & CIPSI(E+PT2)& DMC         \\ \hline
      1 &-281.6729 &-283.0063     & -283.070(1) \\
    239 &-281.7423 &-282.9063     & -283.073(1) \\
 44,539 &-282.0802 &-282.7339     & -283.078(1) \\
541,380 &-282.2951 &-282.6772     & -283.088(1) \\
908,128 &-282.4029 &-282.6775     & -283.089(1)  \\ \hline \hline
\end{tabular}
\caption{Energies in Hartree of C$_2$O$_2$H$_3$N as a function of the number of determinants $N_{det}$ in the trial wavefunction obtained using CIPSI(E), CIPSI(E+PT2) and DMC. CIPSI(E) is the variational energy, while CIPSI(E+PT2) includes perturbative corrections.}
\label{TAB:CIPSI-DMC}
\end{table}

\begin{figure}
\begin{center}
\includegraphics[clip,width=0.85\columnwidth]{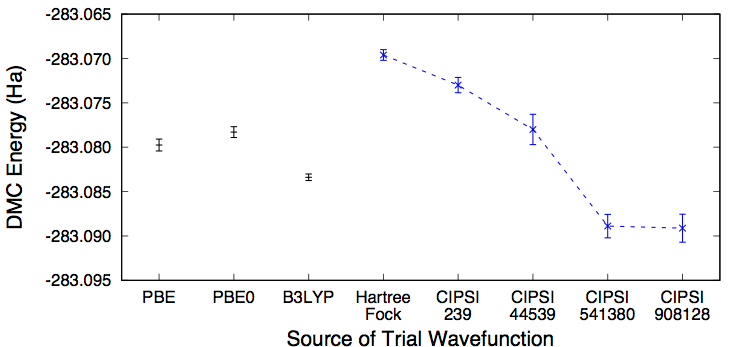}
\end{center}
\caption{DMC energy of C$_2$O$_2$H$_3$N using different trial wavefunctions. Using the aug-ccpVTZ basis, wavefunctions are generated from Hartree-Fock, PBE, and hybrid functionals PBE0 and B3LYP. Multideterminant CIPSI trial wavefunctions contain up to 908128 determinants, as indicated. The dashed line is a guide to the eye and indicates the systematic improvement of the CIPSI wavefunctions.
\label{fig:CIPSI-DMC}
}
\end{figure}

DMC results in table \ref{TAB:CIPSI-DMC} show that the DMC results are converged close to 0.001 Ha, or better than chemical accuracy of 1 Kcal/mol, by around 500000 determiants. Figure \ref{fig:CIPSI-DMC} shows the energy as a function of different single-determinant trial wavefunctions as well as multideterminant wavefunctions generated with CIPSI. The latter show a systematic improvement of the nodal surface as a function of the number of determinants. However, it is also interesting to note that in this case the single determinant B3LYP results are quite accurate, highlighting the importance of orbital selection and optimization to improve efficiency\cite{ClayJCP2015}. 

\section{Utilities}
\subsection{Averaging quantities in the MC data}
\label{sec:averaging} 

\qmcpack includes the \textit{qmca} Python-based tool to average quantities
in the output files and aid in performing statistical analysis. Given
the name of an output file, qmca will compute the
average of each quantity in the file.  Results of separate
simulations can also be aggregated, such as for different twists (twist
averaging), multiple steps (autocorrelation analysis) or multiple
Jastrow parameters (Jastrow optimization).

\begin{figure}[ht!]
\begin{center}
\includegraphics[trim = 0mm 0mm 0mm 0mm, clip,width=0.7\columnwidth]{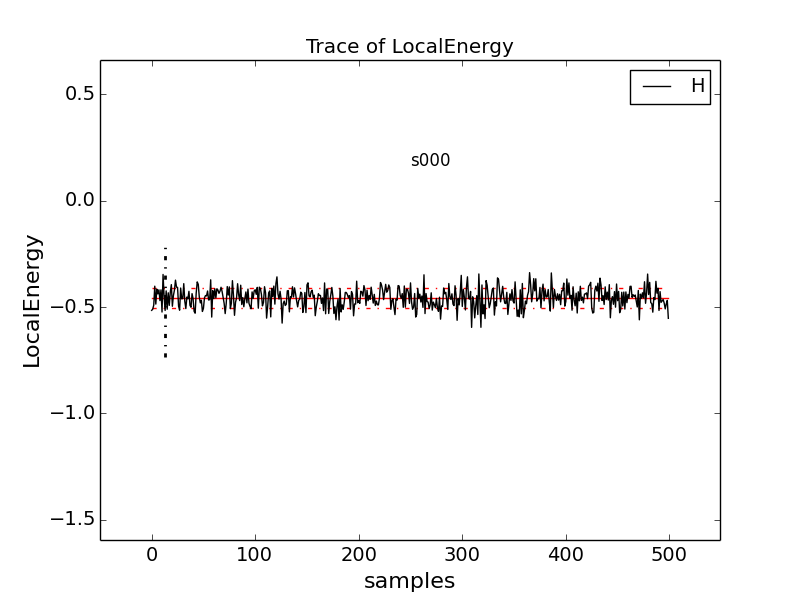}
\end{center}
\caption{Screenshot of qmca trace analysis. The solid black
  line connects the values of the local energy at each Monte Carlo block
(labeled ``samples'').  The average value is marked with a horizontal, solid
red line. One standard deviation above and below the average are marked with
horizontal, dashed red lines. 
The trace of this run is largely centered around the average with no
large-scale oscillations or major shifts, indicating a good quality
Monte Carlo run. 
\label{fig:qmcatrace}
}
\end{figure}

In addition to all the quantities computed by \qmcpack, qmca computes
the data variance and efficiency. qmca also allows visualizing the
evolution of MC quantities over the course of the simulation by a
\textit{trace} offering a quick picture of whether the random walk had
expected behavior as in example figure\;\ref{fig:qmcatrace}.

\subsection{Wavefunction converters}
An important step before running a QMC calculation is to obtain the
trial wavefunction from another electronic structure or quantum
chemical code and convert it into a format readable by \qmcpack. In
addition to the large set of converters available through Nexus,
\qmcpack comes with 2 converters upon compilation. Connections to other codes will be developed on request.

\textit{convert4qmc}: When compiling \qmcpack, an extra binary called
convert4qmc is also created. convert4qmc manages gaussian trial
wavefunctions from codes such as GAMESS\cite{schmidt93}, VSVB\cite{VSVB}
or Quantum Package\cite{QP}. convert4qmc handles the conversion of
single determinant, multideterminants (CASSCF, CI, CIPSI), numerical
and Gaussian basis sets. The output file generated can be either in an
XML or HDF5 format. convert4qmc allows the user to add multiple option
to the wavefunction such as a Jastrow function (2-body, 3-body or
Pade), cusp conditions, or limit the number of determinants to
include.

\textit{pw2qmcpack}: When using a plane wave trial
wavefunction from the PWSCF code in the QE
suite\cite{QuantumEspresso2009,QuantumEspresso2017},  pw2qmcpack.x is used. Source
code patches are included with \qmcpack to produce the pw2qmcpack.x
binary for specific QE versions, necessary to collect and write the
wavefunction in the correct format for \qmcpack. 


\section{Workflow automation using Nexus}
\label{sec:nexus}
Completing the research project path from project conception to
polished results requires a great amount of computational and
researcher effort.  Much of the effort stems from the fact that
obtaining even single, non-production energies from QMC is a
multi-stage process requiring orbital generation (e.g. with a DFT
code), orbital file format conversion, Jastrow optimization via VMC,
subsequent DMC projection, and later analysis.  This process
must usually be repeated many times to ensure convergence of the results with
respect to system size, k-point mesh, B-spline mesh, and DMC timestep,
as well as for the different solids or molecules of interest.
Often this entire process must be performed first in the validation of
pseudopotentials (e.g. via atomic or dimer calculations).
As a further complication, the appropriate computational
environment -- or host computer -- can vary with the stage in the chain from
small clusters for DFT work, mid-size machines for wavefunction
optimization, and sometimes very large supercomputing resources for
DMC or AFQMC.  Simplifying the management of these
processes is of key importance to minimize the full time to solution
for QMC. 

Scientific workflow automation tools have been used with much success 
in the electronic structure community to reduce both the burden on 
the researcher and to reduce the propagation of human error with 
improved systematization.  Packaged with \qmcpack is an automation 
tool, called Nexus \cite{krogel2016_nexus}, which has been tailored 
to the computational 
workflows of QMC.  The system handles several steps 
in the simulation process typically requiring human involvement such 
as atomic structure manipulation, input file and job submission 
script generation, batch job monitoring and error detection, 
selection of optimized wavefunctions, and post-processing of 
statistical data.  Nexus also handles the flow of information between 
simulations in a workflow chain, such as passing on the relaxed 
atomic structure, orbital file information, and optimal Jastrow 
parameters to subsequent simulations that require them.  The system 
is suitable for both exploratory and production QMC calculations 
spanning multiple machines, including those approaching a 
high-throughput style.

Nexus is written in Python following an object-oriented approach to
allow extensibility to multiple simulation codes and host execution
environments.  Nexus currently has interfaces to QE
\cite{QuantumEspresso2009,QuantumEspresso2017}, GAMESS \cite{schmidt93,gordon05}, VASP
\cite{kresse1993,kresse1994,kresse_efficiency_1996,kresse_efficient_1996}, \qmcpack, and a
number of associated post-processing and file conversion tools. Nexus
does not require access to the internet or to an installed database to
run, instead operating only via the filesystem.  Nexus is therefore
suitable for the widest range of computer environments.  Supported
machine environments include standard Linux workstations as well as
high performance computers.  Explicit support exists for
systems at the National Energy Research Scientific Computing Center
(NERSC), the Oak Ridge Leadership Computing Facility (OLCF), the
Argonne Leadership Computing Facility (ALCF), Sandia National Laboratories
high-performance computing resources, the National Center for
Supercomputing Applications (NCSA), the Texas Advanced Computing
Center (TACC), the Center for Computational Innovations (CCI) at
Rensselaer Polytechnic Institute, and the Leibniz Supercomputing
Centre (LRZ).  Variations in the job submission and monitoring
environments at each institution necessitate specific extensions to
ensure operability across this wide range of resources.

Users interact with Nexus by writing short Python scripts that 
generally resemble input files.  Use of such ``input files'' allow 
the user to spend more time on project design rather than execution 
and naturally comprise both a record of calculations performed and 
a means to fully reproduce them.  Nexus has been used extensively in 
summer schools on \qmcpack and recent research papers. An additional
benefit of the workflows is the greatly improved ability to reproduce the
calculations at a later date, and to speed up related research
projects through reuse.


\section{Examples}  
In the following we give examples of recent applications of \qmcpack to illustrate the insights achievable with the currently implemented QMC algorithms and to highlight use of specific features of the application. A molecular example, where the trial wavefunction is systematically converged using CIPSI is given in section\;\ref{sec:cipsi}.

\subsection{Black Phosphorus}\label{example:BP} 
Black phosphorus (BP) has enormous potential for technological
applications because it is layered like graphite and can be exfoliated
to form a 2D material that is naturally a semiconductor. There are
some technical complications in making devices due to effects such as
the degradation of the material when exposed to air or water. Ab
initio calculations can help understand such processes and act as a
laboratory in which to test mitigation strategies. However, the
interaction between the black phosphorus layers was thought to be
dispersive in nature. Non-covalent interactions are a classically
difficult case to treat with density functional based approaches,
while, QMC has been shown to perform as well as the most accurate
quantum chemical methods while also being applicable to extended and
larger systems\cite{ganesh_binding_2014,spanu_nature_2009,benali_vdw}.
We employed \qmcpack to calculate the interaction between black
phosphorus layers as well as the bulk material\cite{shulenburger_bp}.
We found that in one particular arrangement the interaction between
the layers was very well reproduced by one van der Waals-corrected DFT
functional, but as soon as the stacking between the layers was
changed, the functional's performance degraded.

In order to understand this phenomenon, we looked at the change in the
charge density induced by the presence of nearby layers of black
phosphorus. Classically, one might expect very little change in the
charge density due to a van der Waals interaction, and this property
is rigorously upheld by various vdW-corrected DFT functionals.
However, we found a large charge reorganization caused by the presence
of nearby black phosphorus layers and of a character that was
different than that predicted by DFT.

This study took advantage of \qmcpack's ability to handle mixed
boundary conditions (section\;\ref{sec:boundaryconditions}). For BP
monolayers and bilayers, we used periodic boundary conditions parallel
to the layers, but open boundary conditions perpendicular to them.
Also, we found the interaction energies were very sensitive to finite
size effects, and the ability to perform calculation on many different
supercells of the material was crucial to determine their exact form.
The B-spline representation was used for the orbitals in the trial
wavefunctions (section\;\ref{sec:bspline}). Finally, the ability of
the code to evaluate the charge density from the calculation was
crucial.

\subsection{Bilayer \texorpdfstring{$\alpha$}{alpha}-graphyne}
\label{example:graphyne}

Carbon can form two-dimensional sheets of $sp-sp^2$-hybridized atoms,
$\alpha$-graphyne. Its existence was predicted three decades
ago\cite{baughman87} and has recently received a great deal of
attention because of its intriguing potential as a new Dirac
material\cite{zhang11,srinivasu12,malko12}. Among various available
forms of a graphyne-based structures, a bilayer $\alpha$-graphyne
consisting of two stacked two-dimensional hexagonal lattices can be
energetically stabilized in two different stacking modes (AB and Ab
mode) out of a total of six available modes; note that the
$\alpha$-graphyne hexagons are much larger than those of graphene.
While an AB mode has been predicted to exhibit electronic properties
similar to those of an AB bilayer of graphene, Ab-stacked graphyne is
expected to possess split Dirac cones at the Fermi level, and its gaps
can be opened with and applied electric field normal to the
surface\cite{leenaerts13}. Theoretical predictions of these electronic
properties of bilayer graphynes were made by using first-principle DFT
methods\cite{leenaerts13}. However, DFT studies failed to suggest
which of the AB and Ab stacking mode is the most energetically stable
one because of the approximate description of van der Waals forces
within the non-interacting Kohn-Sham scheme. As seen in
Table~\ref{tab:binding}, computed interlayer distances and binding
energies obtained from DFT calculations are strongly scattered
depending on which specific exchange-correlation functional was used.

\begin{table}[t]
\centering
\begin{tabular}{c|cc|cc|c}
  \hline\hline
 \multirow{2}{*}{Method}  & \multicolumn{2}{c|}{$\alpha$-graphyne(AB)} & \multicolumn{2}{c|}{$\alpha$-graphyne(Ab)} & \multirow{2}{*}{$\Delta E_{AB-Ab}$}  \\ \cline{2-5}
  &     $R_{0}$      &      $E_{b}$     &  $R_{0}$         &    $E_{b}$  & \\ \hline                  
DFT-D2 &  3.25      &   13.4      &  3.37    &   13.6     & -0.2  \\ 
vdW-DF &  3.47       &  19.8         &  3.64    &   18.5 &  1.3    \\ 
rVV10  &   3.27      &   17.9     &   3.41     &  17.8   & 0.1  \\ 
DMC  & 3.24(1) & 23.2(2) & 3.43(2) & 22.3(3) & 0.9(4) \\ \hline\hline
\end{tabular}
\caption{Equilibrium interlayer distance $R_0$ (\AA) and binding energies $E_b$ (meV/atom) for an Ab and AB bilayer $\alpha$-graphyne using DMC and various vdW-corrected DFT functionals\cite{shin2017JCTC}. $\Delta E_{AB-Ab}$ indicates the binding energy difference between AB and Ab $\alpha$-graphyne.}
\label{tab:binding}
\end{table}

\begin{figure}[t]
 \includegraphics[width=6.0in]{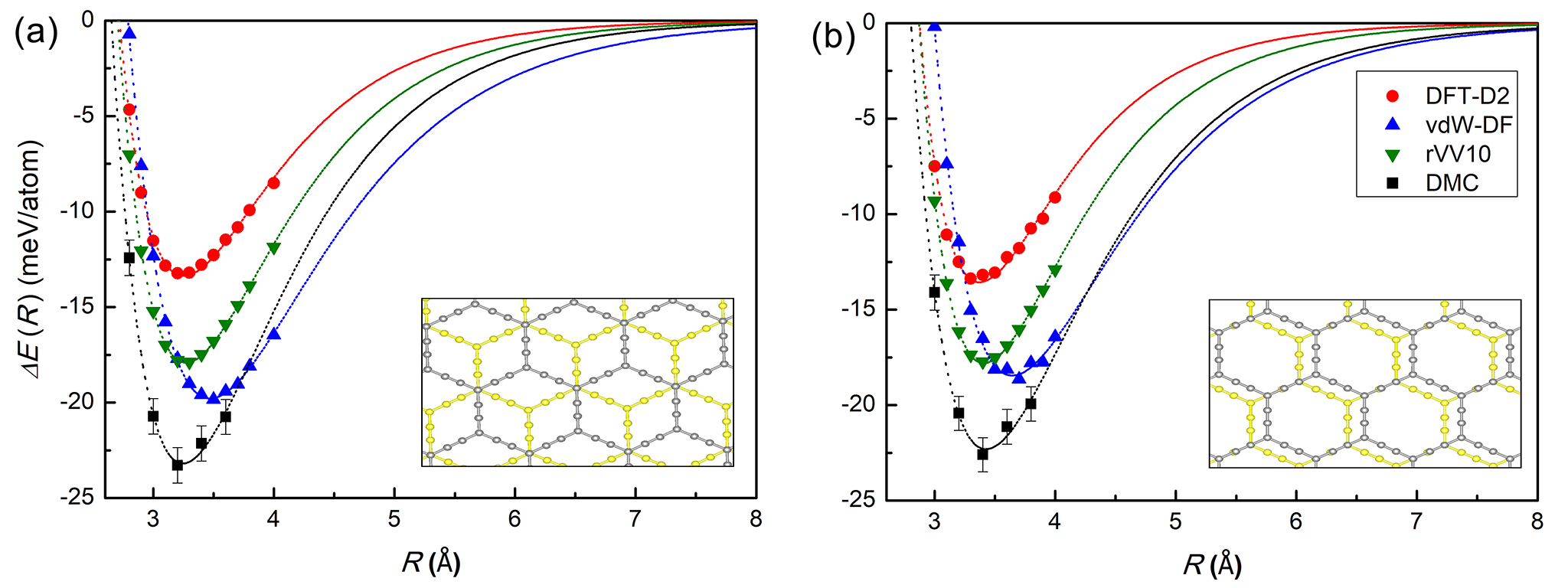}
 \caption{Interlayer binding energy for (a) AB, and (b) Ab stacking mode of a bilayer $\alpha$-graphyne using various DFT XC functionals and DMC as function of interlayer distance $R$. The dotted lines indicate a Morse function fit. Reproduced from \cite{shin2017JCTC}.}
 \label{fig:dftbinding}
\end{figure}

In order to establish which stacking mode is the most stable one, DMC calculations, that include without any approximations the long-range van der Waals interactions, were 
 performed to compute the equilibrium interlayer distance and binding energy for Ab and AB bilayer graphyne\cite{shin2017JCTC}.
DMC equilibrium binding energy for the AB stacking mode (23.2(2)~meV/atom) is estimated to be slightly larger than that for Ab stacking mode (22.3(3)~meV/atom), which suggests that the AB-stacked bilayer is energetically favored over the Ab stacking mode. In comparing the DMC results with results from 
vdW-corrected DFT functionals, including the pairwise correction of Grimme (DFT-D2)\cite{grimme04,grimme06}, self-consistent non-local van der Waals functional (vdW-DF)\cite{dion04}, and simplified non-local vdW-correction (rVV10)\cite{sabatini13}, it was found that those significantly underestimate the interlayer binding energies for both stacking modes (see figure~\ref{fig:dftbinding}).

It is interesting to find which vdW-corrected DFT functional most accurately describe the weak interlayer van der Waals interaction in bilayer $\alpha$-graphynes by comparing the  DFT results to the DMC ones. As Table~\ref{tab:binding} shows, there is no vdW-corrected DFT functional that achieves good accuracy for both interlayer distance and binding energy. However, the rVV10 functional produces almost identical equilibrium interlayer distances as DMC, and the vdW-DF results for interlayer binding energies and $\Delta E_{AB-Ab}$ are the closest ones to the DMC results for both stacking modes. Therefore, one may conclude that rVV10 can provide well-optimized vdW geometries for low-dimensional carbon allotropes, while vdW-DF gives the best vdW energetics among the various vdW-corrected DFT functionals investigated here.  
 
 \subsection{\texorpdfstring{TiO$_2$}{TiO2}}
\label{example:TiO2}
The Row 3 and Row 4 transition metals are extremely useful elements for a number of applications, ranging from magnetic applications to solar cells and catalysis. The reason for their versatility is the partially filled d-shell, which gives rise to a number of oxidation and spin states. The 3d electronic states are also rather localized and give rise to relatively strong electronic correlations, especially in the transition metal oxides, such as NiO and TiO$_2$. This has as a consequence that standard computational approaches using density functional theory within the local density approximation (LDA) or the generalized gradient approximation (GGA) are inadequate and frequently give incorrect results. 

Titanium dioxide (TiO$_2$) occurs in a variety of polymorphs. Three of those occur naturally: rutile, anatase and brookite. Rutile is the most abundant one and is used as a white pigment as well as an opacifier and an ultraviolet radiation absorber, while anatase is the most photocatalytically active polymorph\cite{Diebold2003,HashimotoJapJApplPhys2005}.

In spite of the prevalence and many applications of TiO$_2$, it is very difficult to determine which of the polymorphs is the most stable one. Experimental studies have indicated that rutile is the most stable structure, and anatase and brookite are metastable. However, the enthalpy differences between the polymorphs are very small, of the order of 1~mHa per formula unit, making precise determinations very difficult\cite{Ranade2002,Hanaor2011,Satoh2013,Barnard2004,Barnard2005,Barnard2008,Hummer2009,Hummer2013}. 

Electronic structure calculations using a broad range of local and gradient corrected functionals give anatase as the lowest-energy polymorph, and thereafter brookite and rutile\cite{Muscat2002,Zhu2014}, in supposed disagreement with experiments. DFT+U can give the apparent correct ordering of the polymorphs for a sufficiently large U value\cite{Arroyo-DeDompablo2011,Curnan2015}. Hybrid functionals can also give the correct energetic ordering of the polymorphs, but only using very large fractions of exact exchange\cite{Curnan2015}. 

\begin{figure}[t]
\centering
\includegraphics[width=0.7\columnwidth]{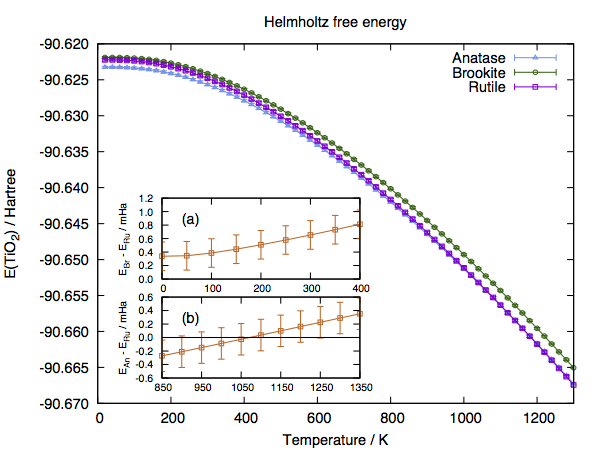}
\caption{The Helmholtz free energy of rutile, anatase and brookite as a function of temperature. All the values are shifted by the calculated Helmholtz free energy of anatase at 0~K. The energy differences between brookite and rutile at 0--400~K, and between anatase and rutile at 450--850~K are provided in the insets (a) and (b). The energy of brookite is always larger than that of the other two solids, while the energy of rutile becomes lower than that of anatase at 650$\pm$150~K. The error bars indicate the statistical uncertainty due to the QMC data used for the 0~K energy differences. Reproduced from \cite{Luo2016NJP}.}
\label{fig:TiO2_free_energy}
\end{figure}

In two recent studies\cite{Luo2016NJP,trail2017prb}, diffusion Monte Carlo calculations were used to examine TiO$_2$ polymorphs with the goal of determining the energetic ordering at zero and finite temperatures. Both studies gave the result that anatase is the most stable polymorph at 0~K, with very small energy differences between rutile and brookite. Finite-temperature Helmholtz free energies were then calculated by including phonon contributions based on DFT phonon calculations. 
The results in both studies were very similar: the finite-temperature contributions from lattice vibrations drove the free energy of the rutile polymorph to be the smallest at temperatures above approximately 650~K (see figure~\ref{fig:TiO2_free_energy} ). These two studies may finally resolve the question of the most stable polymorphs of TiO$_2$: at 0~K, anatase is the most stable one, while lattice vibrations drive a transition from anatase to rutile at about 650~K.

\subsection{The Magn\'eli phase \texorpdfstring{Ti$_4$O$_7$}{Ti4O7}}
Because of its many oxidation states, titanium can form a variety of stoichiometrically different oxides. One particular set are the Magn\'eli phases with the generic formula $Ti_{2n}O_{(2n-1)}$. These form ordered crystals with diminishing band gaps with increasing $n$. In particular, Ti$_4$O$_7$ 
forms a magnetic semiconductor at temperatures below 120~K\cite{Bartho69,Inglis1983}. DFT finds multiple low energy states, but the ordering depends on the functional used.

DMC calculations\cite{Benali2016PCCP} found the same energetic ordering of the lowest three states as the DFT calculations by Zhong {\em et al.}\cite{Zhong2015PRB}, but with larger energy separations.  A detailed examination of the spin densities 
and local moments on the Ti$^{4+}$ ions showed that the DFT methods actually gave very good representations of the total moments. However, the orbital alignments were different in DMC, especially in the FM state. This certainly will give rise to different energy differences compared to DFT as the energy differences between the states can rather accurately be attributed to Heisenberg exchange between the magnetic Ti ions, which will differ with the different orbital alignments.

\subsection{\texorpdfstring{VO$_2$}{VO2}}
\label{example:VO2}
Vanadium dioxide (VO$_2$) has a rich phase diagram which can be exploited in novel device applications. At ambient pressure, the unstrained VO$_2$ undergoes a metal to insulator transition (MIT) at $T_c\approx 341$ K from the low temperature monoclinic M1 phase to the high temperature rutile (R) phase \cite{PhysRevLett.3.34,NanoLett2010_VO2_Phases,HuberNanoLett2016}. This phase transition has been studied rather extensively, in part due to the on-going debate on the driving mechanism of this MIT. The challenge in describing this material is often related to the representation of electron correlations in this strongly correlated electron material \cite{RevModPhys.70.1039,PhysRevApplied.7.034008,PhysRevB.11.4383,Eyert_AnnPhys,PhysRevB.86.075149,PhysRevB.95.035113,Budai_Nature}. Commonly, experimental studies are accompanied with density functional theory (DFT) calculations for better understanding. Additionally, DFT has been used in isolation to provide insight into the mechanism of the MIT \cite{LeeScience371,ChenNanoLett2017,AppavooNanoLett2014,WANG2014126}. However, the failings of DFT in this context are also well known \cite{zheng2015,PhysRevB.86.081101,PhysRevB.90.085134}.  If functional development led to systematic improvement\cite{medvedev_density_2017}, this should be measurable in both the total energy and the electron density; the two properties that the exact functional must perfectly reproduce.

\begin{figure*}[t]
\includegraphics[width=\textwidth]{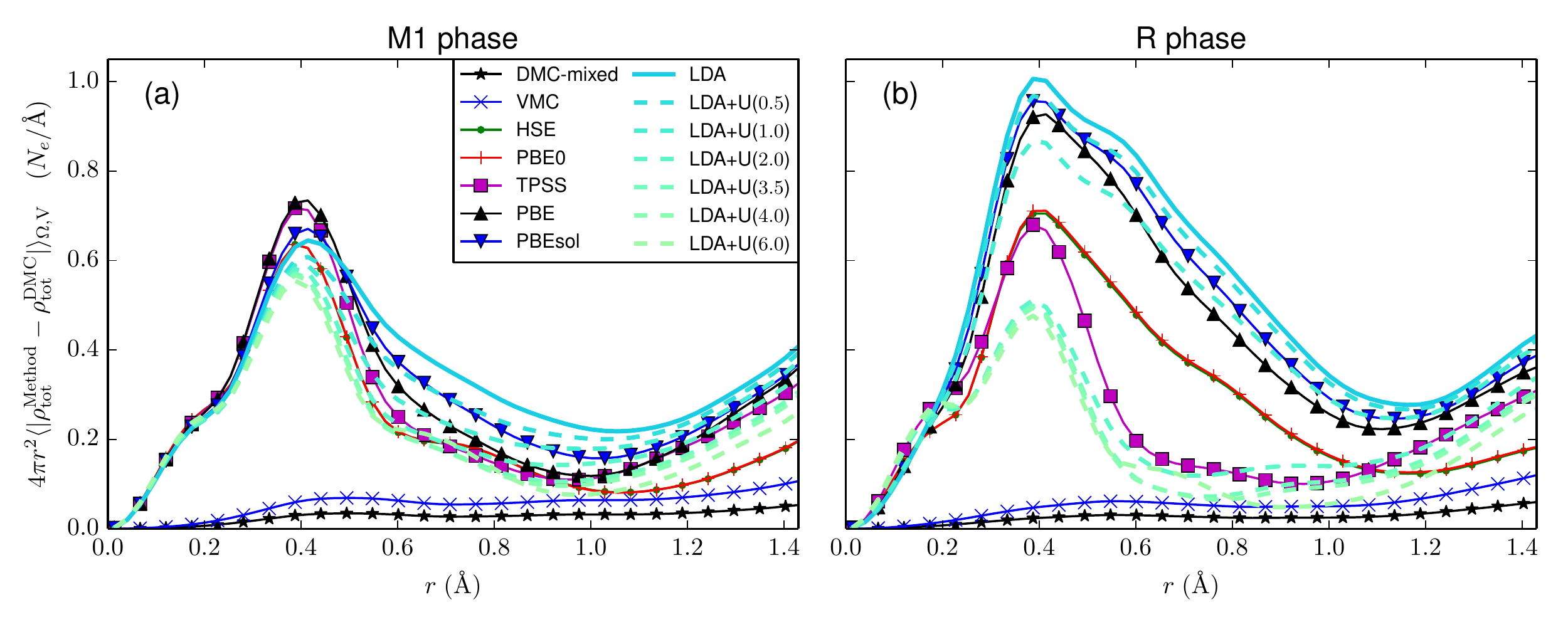}
\caption{\label{VO2density}Radial distribution function of absolute total density difference 
from extrapolated estimator around V atom for various theoretical methods 
using RRKJ pseudopotentials: (a) M1 phase, and (b) 
R phase. As a function of increasing U value the LDA+U density tends 
to an improved accuracy, i.e., as U is increased LDA+U curves get 
closer to zero. Reproduced from \cite{Kylanpaa2017PRM}.}
\end{figure*}

Recent QMC calculations of VO$_2$ were used as a reference in assessing various DFT formulations\cite{Kylanpaa2017PRM}. Supercells of 48 atoms were used to model the antiferromagnetic ground state of the M1 and R phases, and a $3\times 3\times 3$ grid was used in twist-averaging. The QMC calculations used LDA+U orbitals for the nodes, where the optimized U-value of 3.5 eV was obtained from DMC as the value that minimized the variational DMC energy.

In general, it was observed that the best description of energetics between the structural phases did not correspond to the best accuracy in the charge density. An accurate spin density was found to lead to a correct energetic ordering of the different magnetic states. However, local, semilocal, and meta-GGA functionals tend to erroneously favor demagnetization of the vanadium sites, which can be reconciled in terms of the self-interaction error. The metrics used also revealed the limitations in the description of correlated 3d-orbital physics present in currently available functionals. This is evident, e.g., in the density metrics shown in figure~\ref{VO2density}, where spatial variations in the electron density with respect to DMC reference are shown; the extrapolated estimator is used as the DMC reference.

\subsection{NiO and potassium-doped NiO}
\label{example:K-NiO}
Nickel oxide is the poster-child for correlated transition metal oxides. It has a simple rocksalt structure and is at low temperatures a Type II antiferromagnetic insulator. Not only is NiO interesting as a prototypical correlated transition metal oxide, but it is typically a $p$-type semiconductor, one of a very small number of $p$-type oxides, because as-deposited NiO typically has a Ni deficiency. Stoichiometric NiO can also be p-doped with suitable monovalent elements, {\em e.g.,} with Li or K. 
This brings the basic questions of what are the energetics of vacancies or substitutional dopants in NiO, and how do their descriptions in DFT-based calculations differ from those based on QMC?

DMC calculations\cite{shin2017PRM} were performed on 64-atom supercells with a single substitutional K dopant or Ni or O vacancies.
DMC results for ground state properties were in very good agreement
with experiments, compared to the DFT results (see
Table~\ref{tab:DMC_bulk_NiO}). There was also a large difference in
defect creation energies, with DFT energies underestimating them
(Table \ref{tab:K_formation_NiO}). The optical gaps were also underestimated by DFT, as expected. DMC calculations of the optical gap were larger, in fact much larger than experimental values. Calculations of the gap using a 128-atom supercell reduced the gap, but the variance was large enough for this size cell that the reduction of the gap for the larger supercell was not statistically significant. In any case, the gap calculations indicate that finite-size corrections can be significant in DMC gap calculations and also point to the need for improved excited state methodologies in QMC calculations of solids, where the errors are currently larger than for ground states.

\begin{table}[t]
\centering
\caption{
Values of lattice constant ($a$), bulk modulus ($B_0$), and cohesive energy ($E_{coh}$) for AFM-II type NiO obtained from a Vinet fit of the equation of state computed using GGA+U and DMC at U=U$_{opt}$ and a 16-atom type II AFM NiO supercell.}
\label{tab:DMC_bulk_NiO}
\begin{tabular}{cccc}
\hline \hline
     method        &      $a$ (\AA)     &  $B_{0}$ (GPa) & $E_{coh}$ (eV/f.u.) \\ \hline
 GGA+U &     4.234     &   192     &   8.54      \\ 
DMC &  4.161(7)    &    218(14)   &   9.54(5)     \\ 
Experiment$^1$   &   4.17  &  145-206 & 9.5  \\ \hline \hline                      
\end{tabular}
\begin{flushleft}
$^1$\cite{Lide95}.\\
\end{flushleft}
\end{table}

\begin{table}[t]
\centering
\caption{DMC formation energies of a K dopant ($E_{f}$) NiO under O-rich condition, and optical gap ($E_{g}$) for NiO and K-doped NiO in a 64-atom supercell\cite{shin2017PRM}. Energies are in eV.}
\label{tab:K_formation_NiO}
\begin{tabular}{cccc}
\hline \hline
     method        &      $E_{f}$(K) & $E_{g}$ (NiO) & $E_{g}$ (KNiO)  \\ \hline
 GGA &  1.9 &  1.4 &  0.7 \\
GGA+U & 0.6 &  3.6 &  2.9 \\ 
\multirow{2}{*}{DMC} &  \multirow{2}{*}{1.3(3)} & 5.8(3) & \multirow{2}{*}{4.8(4)}    \\ 
     &           &  5.0(7)$^{1}$     &     \\
Exp.&       -   &  4.3$23$  &  3.7 - 3.9$^3$ \\ \hline \hline
\end{tabular}
\begin{flushleft}
$^1$Optical gap calculated in a 128 atoms supercell. \\
$^2$Reference~\cite{sawatzky84,hufner84,hufner86,zaanen86}.\\
$^3$Reference\cite{yang11-2}.
\end{flushleft}
\end{table}

\subsection{Pseudopotential development and testing}
\label{example:pp_testing}

Testing pseudopotentials is an important part of QMC due to (i) the historic challenge of developing accurate pseudopotentials for many-body methods such as QMC and historical reuse of DFT or Hartree-Fock-derived potentials, and (ii) the importance of checking any biases unique to QMC such as the T-moves and locality approximations in DMC.  Comparing atomic properties such as ionization potentials, and dimer properties such as bond lengths and binding energies obtained by DMC with experimental or quantum chemical results, provides a robust, inexpensive, and transferable test of pseudopotential quality.  Recently, \qmcpack and Nexus \cite{krogel2016_nexus} were used jointly to validate a collection of newly developed pseudopotentials for the early-row transition metals (Sc-Zn) \cite{krogel2016_pp}.

DMC calculations were performed for five atomic 
charge states (ranging from neutral to 4+) and at nine transition 
metal-oxygen dimer bond lengths for each species.  Orbitals were 
generated with QE using the experimental spin 
multiplicity within LDA \cite{ceperley1980,perdew1981} or HSE 
\cite{heyd2003,heyd2006} for atoms and LDA for dimers.  Two-body 
Jastrow functions were optimized at each atomic charge state and at the 
equilibrium geometry for dimers.  Subsequent DMC calculations were 
performed with both T-moves and the locality approximation to assess 
the affect of pseudopotential localization errors.  This large number of calculations is best handled using workflow software such as Nexus.

The resulting DMC atomic ionization potentials and dimer bond 
lengths, binding energies and vibration frequencies were compared 
with prior DMC results using Gaussian-based Hartree-Fock 
\cite{hartree1928,fock1930,roothaan51}
pseudopotentials, all electron quantum chemistry results using MRCI 
\cite{werner1988,knowles1988} and CCSD(T) 
\cite{raghavachari1989,bartlett1990,knowles1993} approaches, and 
experiment.  On essential all measures,
the various theoretical approaches performed similarly well compared 
with experiment.  The current pseudopotential DMC results were within 
0.2 eV of experiment on average for atomic ionization potentials and 
dimer binding energies, with the T-moves and locality approximation 
generally agreeing to 0.05 eV for these energy differences.  
Equilibrium bond lengths were found to be within 0.5\% of the 
experimental values, while the more sensitive vibration frequencies 
agreed to around 3\%.  This work as well as subsequent studies have 
verified the quality of the new potentials for QMC studies of 
transition metal containing systems, but further improvements are desirable (section\;\ref{sec:future_pp}).


\section{Future extensions and challenges}

\subsection{Pseudopotentials}
\label{sec:future_pp}

The atomic core energies scale quadratically with
the nuclear charge, $Z$, while the valence energies stay essentially
constant across all chemical elements.  The energy fluctuations from
the core therefore dominate any QMC calculation and, indeed, the
overall cost is $\propto Z^a$ with $a$ between 5.5 and 6.5\cite{CeperleyJSP1986,HammondJCP1987}.  It is
therefore highly desirable to construct an effective, valence-only
Hamiltonian with the atomic cores removed.  In fact, a related problem
is encountered also in one-particle calculations with plane waves.
For that matter, even in many-body calculations with heavy atoms fully
correlating the core(s) would eventually dominate the calculations
regardless of the employed method. 

Pseudopotentials (PPs) and the closely related effective core
potentials (ECPs) have been used in condensed matter calculations as
well as in basis set quantum chemical calculations for several
decades\cite{martin_electronic_2004}. At the very basic level, the
atomic cores and corresponding degrees of freedom are replaced by
PPs/ECPs operators that mimic the action of the core states on the
valence electrons. Traditional PP/ECP constructions are based on
one-particle solutions of the atom and typically involve norm
conservation/shape consistency so that the pseudo-orbitals match the
true original all-electron orbitals outside some appropriate core
radius. Many advances for ECPs/PPs have been proposed that generalize,
make more efficient, improve or more accurately reproduce the true
atomic properties, e.g. Refs.
\cite{bachelet_pseudopotentials_1982,bylander_exact_1992,vanderbilt_soft_1990,hamann_optimized_2013}. For
QMC calculations, one improvement that has been used to improve the
transferability has been based on fitting HF (Hartree-Fock) or DF
(Dirac-Fock) energy differences for a set of atomic and ionic
excitations. Furthermore, many-body constructions have been suggested
by means of reproducing the (correlated) one-particle density matrices
beyond a certain radius or by improving upon DFT solution for a given
atom using many-body perturbation theory.

Despite all these elaborate and sophisticated efforts, at present,
pseudopotentials remain a very laborious part of QMC studies.  The
main reason is that QMC often reaches accuracy {\em beyond} and
sometimes {\em well beyond} the accuracy of traditional and even
currently most advanced PP/ECP constructions.  Additional
complications come from dealing with the non-locality in the QMC
framework that introduces further complications and demands on quality
of the trial wavefunctions. As a result, PP/ECP for every element has
to be painstakingly retested and benchmarked anew and if the
inaccuracies sufficiently bias the valence properties of interest, one
has to go back to square one and construct a more accurate
PP/ECP. Furthermore, techniques in DFT that use solid state
results to improve transferability,
e.g. \cite{schlipf_optimization_2015}, are not practical in QMC
because of the computational cost of the large supercells required to
converge finite size errors.

To overcome these highly technical but important barriers it
is desirable to develop a new generation of PPs/ECPs based
on correlated constructions from the outset, provide high accuracy
that would not limit the subsequent QMC results, and also be 
efficiently used in QMC.  In addition, one would like to have
flexibility in choosing the core-valence partitioning and
transferability not only at ambient (equilibrium) conditions but also
at high pressures, non-equilibrium conformations and other broader set
of conditions.

The goal is therefore not only to reproduce the properties of the atom
within one-particle theory and then hope for best in many-body
calculation. Our effort would be focused on reproducing the true
many-body properties of the original system(s), i.e., atom(s), in
a variety of settings.  For this purpose, we plan to derive new
generation of PPs/ECPs that would be based on a number of new criteria
targeted to uphold its accuracy and fidelity to the true original
many-body Hamiltonian. In particular,

\begin{enumerate}
\item  we plan to construct an atomic ECP operator that will match a
subset of {\em the many-body spectrum as close as possible to the
original atomic Hamiltonian.} This will involve a set of states that
have the largest weights in molecular, surface, solid or other
chemical settings;

\item include more options for multiple core-valence partitioning
whenever appropriate;

\item express PPs/ECPs in a simple representations/forms that enable
their use with multiple methods ranging from traditional DFT and plane
wave-based packages to many-body approaches based on stochastic and
explicit expansions in basis set methods;

\item try to capture all of the relevant physics that is feasible at the
current state-of-the-art, e.g., impact of {\em correlated cores together
with correlated valence}, describe relativity with the best available
account of correlation, explicit treatment of spin-orbit effects, etc.
\end{enumerate}

Clearly this is an ambitious plan that will require significant effort
and time, and as such it is almost a never ending task (since it is
almost always possible to slightly improve upon the previous version).
Nevertheless, we believe that equally important is the adoption of
{\em new standards}: many-body instead of one-particle framework,
testing and benchmarking by a multitude of methods that cross-validate
the quality of the PPs/ECPs, and systematic documentation and
improvements so that PPs/ECPs can be used without {\em endless
retesting} and {\em with true many-body quality of the corresponding
operators shown upfront.} To-date, explicitly correlated ECPs have been
developed for a selection of first and second row elements
\cite{bennettmelton17}, with future developments for the rest of the
periodic table underway.

These developments will be collected on a new community website,
\url{http://pseudopotentiallibrary.org}, that will be used for storing
the data and will also include tests and benchmarks. The PPs/ECPs will
also be available as a part of the \qmcpack package.

\subsection{Spin-orbit interaction}
  
Until now, almost all electronic structure QMC calculations have been
carried out with spins of individual electrons being fixed, up or
down. This is easy to justify for Hamiltonians without explicit spin
operators. Such Hamiltonians commute with any spin, so that up or down
orientations are conserved and the solution of the many-body problem
is therefore confined to the space of spatial coordinates, and spin
states are imposed as a symmetry (e.g. a triplet state).  Many
interesting phenomena involve the interaction between the spin and
spatial degrees of freedom, such as the spin-orbit
interaction. Recently, dynamic spins as quantum variables has been
realized in DMC in order to treat such interactions directly, see
\cite{melton-pra} and \cite{melton-jcp}.  We plan to implement dynamic
spins into \qmcpack, and a brief outline of the necessary changes are
outlined below.

In order to implement spin-dependent operators, spin variables need to be introduced as dynamic quantum variables.
This requires a change from one-particle orbitals to one-particle spinors
\begin{equation}
    \chi_n(\mathbf{r}_i,s_i) = \alpha \varphi^\uparrow_n(\mathbf{r}_i)\chi^\uparrow(s_i) + \beta \varphi^\downarrow_n(\mathbf{r}_i)\chi^\downarrow(s_i)
    \label{eqn:spinor}
\end{equation}
where the spatial orbitals can be different for the spin up and down channels. 
The spin functions $\chi^{\uparrow(\downarrow)}$ take on discrete values
in the minimal spin representation, namely 
$\chi^\uparrow(1/2) = \chi^\downarrow(-1/2) = 1$ and $\chi^\uparrow(-1/2) = \chi^\downarrow(1/2) = 0$. 
The simplest antisymmetric wavefunction is then a single determinant of the one-particle spinors, 
rather than the product of spin-up and
spin-down determinants. In order to increase the efficiency of sampling the spin degrees of freedom, 
the spinors will be represented using a continuous (overcomplete) representation 
\begin{equation}
    \chi^{\uparrow(\downarrow)}(s_i) = e^{\pm i s_i}
    \label{eqn:spin_function}
\end{equation}
where $s_i \in (0,2\pi)$. These explicitly varying spins result in complex wavefunctions, 
for which we must abandon the fixed-node approximation 
that applies to real-valued wavefunctions. By writing the many-body wavefunction in terms 
of an amplitude and phase $\Psi = \rho \exp\left[ i\Phi \right]$, 
the real part of the Schr\"{o}dinger equation becomes
\begin{equation}
    -\frac{\partial \rho}{\partial \tau} = \left[ -\frac{1}{2}\nabla^2 + V + \frac{1}{2}|\nabla \Phi|^2 + W^{Re}\right]\rho
\end{equation}
where $W^{Re}$ is $\textrm{Re}\left[ \Psi^{-1} W \Psi \right]$ and $W$
is any nonlocal operator such as a pseudopotential.  Since we do not
know the exact phase, we approximate it by the use of a trial phase
using the Fixed-Phase approximation \cite{ortiz-fp93}.  Since $\rho$
is positive-definite, there is no nodal surface and the DMC algorithm
is seemingly the same as for bosonic ground states with an additional
potential provided by the trial phase.

Sampling of the spin variables can be achieved by introducing a spin ``kinetic energy'', namely
\begin{equation}
    T_{s_i} = -\frac{1}{2\mu_s} \left[ \frac{\partial^2}{\partial s_i^2} + 1 \right]
    \label{eqn:spin_kinetic}
\end{equation}
such that it annihilates an arbitrary spinor and does not contribute
to the total energy.  However, the introduction of this operator
modifies the DMC Green's function to include a diffusion and drift
term for the spin variables $\mathbf{S} = (s_1,s_2,\ldots,s_N)$.  The
$\mu_s$ is a spin mass, which can be interpreted as a time step for
the spin variable propagation.

Once the spins are treated as variables rather than static labels,
spin-dependent Hamiltonians can be treated.  For the spin-orbit
integration, a generalization of the non-local operators used in QMC
calculations is necessary. Relativistic quantum chemistry calculations
utilize the semi-local form
\begin{equation}
    W_i = \sum\limits_\ell  \sum\limits_{j=-|\ell-1/2|}^{\ell+1/2} \sum\limits_{m=-j}^{j}W_{\ell j}(r_i) | \ell j m_j \rangle \langle \ell j m_j |
    \label{eqn:rel_ecp}
\end{equation}
for fully relativistic pseudopotentials that include
scalar-relativistic and spin-orbit effects
\cite{relativistic-ecp81}. Utilizing the spin representation
introduced above, the contribution from the non-local pseudopotential
is inherently complex.  When evaluating the localization of the
pseudopotential with the trial wavefunction
$\textrm{Re}\left[\Psi_T^{-1}W\Psi_T\right]$, one will encounter terms
such as $\langle \Omega s | \ell j m \rangle$, where $\Omega$ is the
solid-angle and $s$ is the spin variable for an individual electron.
These terms are given as
\begin{eqnarray}
    \langle \Omega s | \ell, j = \ell\pm1/2, m_j \rangle = \pm \sqrt{\frac{\ell\pm m + 1/2}{2\ell+1}}Y_{\ell,m-1/2}(\Omega)e^{is} \nonumber \\
    +\sqrt{\frac{\ell \mp m + 1/2}{2\ell + 1}}Y_{\ell,m+1/2}(\Omega)e^{-is}
    \label{eqn:ecp_eval}
\end{eqnarray}
By summing over each electron and ion and taking the real part, one
obtains the contribution to the local energy from the complex
non-local pseudopotential in the locality approximation.  As is
described in \cite{melton-jcp}, a generalization to the Casula T-moves
\cite{Casula2006,Casula2010} has been constructed for complex
non-local operators such as the relativistic pseudopotentials above,
and will also be included.

\subsection{Adapting to trends in High Performance Computing}
  
Predicting details about future high performance computing architectures is
difficult, but some general trends can be observed that should motivate
application design decisions in an effort of increasing
forward portability. As
Moore's Law and Dennard scaling slow due to physical limitations, recent
strategies for gaining performance have largely consisted of increasing the
parallel capability of hardware. However, the time and power costs of moving
data during program execution also limits the traditional increase in
parallelism by scaling out commodity-type nodes and connecting them by a high
performance interconnect. As a result, parallelism is being added both at the
overall system level as before, and also within the processors by increasing
thread count, and within the node by adding specialized compute accelerators such
as GPUs, FPGAs, etc. Besides increasingly complex compute hierarchies, memory is
also becoming hierarchical and more complex, e.g. with the addition of
non-volatile memory and various types of on-package high-bandwidth memories.
Although the traditional memory structure already involved multi-level caches 
and system RAM, this complexity was almost entirely managed by the hardware
controllers and compilers.  However, as the memory hierarchy deepens, it could
become necessary for the application developer to more actively manage its usage
to realize desired performance increases. Adapting to these changes is
very important for QMC because the high computational cost often
results in the methods being run on the latest supercomputers with the
newest architectures.

While QMC methods have considerable natural parallelism that makes
maintaining parallel efficiency fairly straightforward on most traditional
hardware, the real desire is to reduce the overall time to solution for any
given problem of interest.  This means that simply mapping more walkers
 to more available threads does not scale indefinitely
for our purposes, and a multifaceted strategy is needed to leverage increasingly
complex modern computer architectures.  The emergence of GPU-based HPC
architectures around half a decade ago indicated that maintaining high 
performance was no longer ``business as usual.'' 

In order to take advantage of the massive increase in the number of lightweight
threads that were offered by GPUs, the parallel strategy in \qmcpack was
re-mapped from walker-based parallelism to particle-based parallelism.  However,
as compute accelerators get more powerful with an increasing number of available
threads, and more powerful
threads have their own vectorization capabilities, additional parallelism
must be exploited. For example, so that multiple threads can be assigned to a single particle
within each walker when working on compute intensive operations such as the 
inverse matrix update and Jastrow function calculations. 
Meanwhile, the natural 
parallelism present in the Monte Carlo method cannot be abandoned, as it will 
always be desirable to use more walkers. For this piece of the strategy to 
continue to be viable, better protocols for equilibration are needed to ensure 
that the overall time to solution can continue to be
reduced. 

It is common for HPC application developers to strive for a forward portable
design to minimize the effort required in adapting their code to new
architectures. With the trend for increasing complexity and a variety of compute
and memory architecture configurations, a flexible application design is as
important as ever for attaining good ``performance portability'' - the ability
to run efficiently on significantly different architectures,
e.g. conventional processors, GPUs, and potentially even FPGAs.

\qmcpack is being actively developed to adopt to the changing
computational landscape. In particular, to (i) run efficiently on
systems with fewer ``fat'' nodes with deeper heterogeneous compute
capabilities as well as systems with a larger number of more
homogeneous ``thin'' nodes by supporting multiple granularities of
computation, (ii) accommodating multiple types of fine-grained
multi-threading and vectorization to fully utilize the processors and
compute accelerators, and (iii) make effective use of the increasingly
complex memory hierarchy. To facilitate this task, simplified
``miniapps'' have been developed for real space QMC and AFQMC. These
are distributed via \url{https://github.com/QMCPACK}. The miniapps
encapsulate the main operations of each method while avoiding the full
complexity of the main application and enabling use as testbeds for
different programming models and middleware layers that help treat the
above complexities. In the long term, we hope for a future C++ standard
that provides a migration path to an effective and portable solution.


\section{Summary}

We have described the current capabilities of the open source and
openly developed \qmcpack Quantum Monte Carlo package. The methods
implemented enable full many-body electronic structure calculations to
be performed for a wide range of molecular to periodic
condensed matter systems, including metals, and using either
all-electrons or pseudopotentials in the Hamiltonian. By solving the
Schr\"odinger equation using statistical methods, large and complex
systems can be studied to unprecedented accuracy, including systems
where other electronic structure methods have difficulty.

\qmcpack contains both real space and orbital space Quantum Monte Carlo
algorithms. Both classes of algorithm involve only limited and
well-controlled approximations and can potentially be systematically
converged to give near exact results. By virtue of the different
approximations and convergence routes involved, these enable
cross-validation between the methods and promise to significantly
strengthen predictions where very different methodologies agree. The
methods are well suited to today's supercomputers and the
architectural trends towards exascale computing. Both the parallel
scalability and on node numerical performance of \qmcpack are state of
the art, minimizing time to scientific solution.  Updates are planned
to take full advantage of the trends in high-performance computing in
a performance-portable manner. Due to the rapid development in the
fundamental Quantum Monte Carlo algorithms and methodology, we plan to
continue to extend \qmcpack to incorporate the best new methods from
the current authors as well as those developed or contributed by the
wider Quantum Monte Carlo community.

\section{Acknowledgments}

Major support for \qmcpack is currently provided by the
U.S. Department of Energy, Office of Science, Basic Energy Sciences,
Materials Sciences and Engineering Division, as part of the
Computational Materials Sciences Program and Center for Predictive
Simulation of Functional Materials. Exascale and performance
portability efforts (GL, MD) were supported by the Exascale Computing
Project (17-SC-20-SC), a collaborative effort of the U.S. Department
of Energy Office of Science and the National Nuclear Security
Administration. Previous support has included: the Predictive Theory
and Modeling for Materials and Chemical Science program by the
U.S. Department of Energy Office of Science, Basic Energy Sciences;
``QMC Endstation'' supported by Accelerating Delivery of Petascale
Computing Environment at the DOE Leadership Computing Facility at
ORNL; Argonne Leadership Computing Facility, which is a
U.S. Department of Energy Office of Science User Facility operated
under contract DE-AC02-06CH11357; Oak Ridge Leadership Computing
Facility, which is a DOE Office of Science User Facility supported
under Contract DE-AC05-00OR22725; the Lawrence Berkeley National
Laboratory Chemical Sciences Division under U.S. Department of Energy
Office of Science contract DE-AC02-05CH11231; ``PetaApps'' supported
by the U.S.  National Science Foundation; and the Materials
Computational Center, supported by the U.S. National Science
Foundation.  Sandia National Laboratories is a multimission laboratory
managed and operated by National Technology \& Engineering Solutions of
Sandia, LLC, a wholly owned subsidiary of Honeywell International
Inc., for the U.S. Department of Energy’s National Nuclear Security
Administration under contract DE-NA0003525.


\section*{References}
\providecommand{\newblock}{}

\end{document}